\documentclass[
reprint,
superscriptaddress,
amsmath,
amssymb,
pre,
]{revtex4-1}

\usepackage{graphicx}
\usepackage[utf8]{inputenc}
\usepackage[english]{babel}
\usepackage[colorlinks=true,allcolors=blue]{hyperref}

\newcommand{\ERN}{\lvert E_\mathrm{RN} \rvert}
\newcommand{\ERP}{\lvert E_\mathrm{RP} \rvert}

\begin{document}
\title{Protein driven lipid domain nucleation in biological membranes}
\doi{10.1103/PhysRevE.100.042410}
\author{Moritz Hoferer} 
\affiliation{ETH Zurich, Z\"urichbergstrasse 18, 8092 Zurich, Switzerland}
\author{Silvia Bonfanti} 
\affiliation{Center for Complexity and Biosystems, Department of Physics, 
University of Milan, Via Celoria 26, 20133 Milan, Italy}
\author{Alessandro Taloni} 
\affiliation{CNR - Consiglio Nazionale delle Ricerche, Istituto dei Sistemi Complessi,
Via dei Taurini 19, 00185 Rome, Italy}
\author{Caterina A. M. La Porta}
\affiliation{Center for Complexity and Biosystems, Department of Environmental Science and Policy,
University of Milan, Via Celoria 16, 20133 Milan, Italy}
\affiliation{CNR - Consiglio Nazionale delle Ricerche, Istituto di Biofisica, 
Via Celoria 16, 20133 Milan, Italy}
\author{Stefano Zapperi}
\email{stefano.zapperi@unimi.it}
\affiliation{Center for Complexity and Biosystems, Department of Physics, 
University of Milan, Via Celoria 26, 20133 Milan, Italy}
\affiliation{CNR - Consiglio Nazionale delle Ricerche, Istituto di Chimica della Materia Condensata e di Tecnologie per l'Energia, 
Via R. Cozzi 53, 20125 Milan, Italy}
\date{\today}

\begin{abstract}
Lipid rafts are heterogeneous dynamic lipid domains of the cell membranes that are involved in several biological processes, like protein and lipid specific transport and signaling. 
Our understanding of lipid raft formation is still limited, due to the transient and elusive nature of these domains {\it in vivo}, in contrast to the stable phase-separated
domains observed in artificial membranes. Inspired by experimental findings highlighting the relevance of transmembrane proteins for lipid rafts, we investigate lipid domain nucleation by coarse-grained molecular dynamics and Ising model simulations. We find that the presence of a transmembrane protein can trigger lipid domain nucleation in a flat membrane from an otherwise mixed lipid phase. Furthermore, we study the role of the lipid domain in the diffusion of the protein showing that its mobility is hindered by the presence of the raft.
The results of our coarse-grained molecular dynamics and Ising model simulations thus coherently support the important role played by transmembrane proteins in lipid domain formation and stability.
\end{abstract}

\maketitle
\section{Introduction}
Biological membranes are ubiquitous cell structures 
that play crucial roles for the regulation of many vital processes, such as selective permeability and homeostasis maintenance, signaling and ion conductivity~\cite{andersen2007bilayer,Ramakrishnan2014}. Membranes display a huge complexity in their microscopic structure being composed by more than one thousand different lipids,
cholesterol, several types of proteins and a small number of carbohydrates.  
In an aqueous solvent, lipids spontaneously arrange in a double layer (with hydrophilic heads pointing outside and hydrophobic tails inside the sheet) that can forge different structures depending on the environment, such as flat membrane or spherical and cylindrical micelles. 

\begin{figure}[b]
	\centering
	\includegraphics[scale=1]{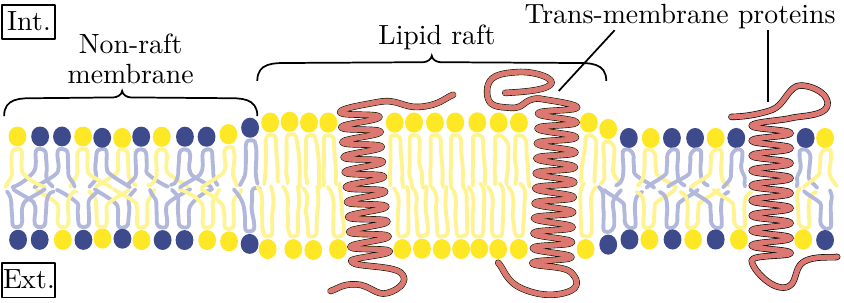}
	\caption{
		\label{fig:figure1}
		\textbf{Sketch of lipid raft domain in a cell membrane.} Raft regions are composed of raft lipids (yellow) and transmembrane proteins (red). Non-raft regions of the membrane are characterized by non-raft-lipids (blue), raft lipids and proteins in a mixed phase. ``Int." (``Ext.'') indicates the intracellular (extracellular) space. The figure is adapted from~\cite{wikimedia}; CC BY-SA 2.5 license.
	}
\end{figure}

An intriguing feature of membranes is the emergence of lipid rafts, highly dynamics lateral heterogeneous domains (in mobility as well as formation and disruption)~\cite{sezgin2017mystery} at different length scales ($\sim \mathrm{nm}$) and timescales ($\sim \mathrm{ms}$)~\cite{brown1998functions,anderson2002role}.
Rafts consist in aggregates of saturated lipids and proteins such as glycosylphosphatidylinositol(GPI)-anchored or transmembrane proteins (see Fig.~\ref{fig:figure1}),
induced by protein-protein and protein-lipid interactions and are thought to be responsible for many membrane-associated functions~\cite{simons1997functional,lingwood2010lipid,harder1997caveolae,varma1998gpi,brown1998sphingolipid,
brown2000structure,douglass2005single,pike2006rafts,lingwood2010lipid,owen2012sub,schulz2012beyond,honigmann2014scanning,lorent2017structural}. 
Recent advances in experimental detection techniques~\cite{sezgin2011fluorescence}, such as fluorescence microscopy~\cite{Klymchenko2014},
super resolution optical microscopy~\cite{eggeling2015super}, interferometric scattering microscopy~\cite{ortega2012interferometric}, single particle tracking~\cite{kusumi2005paradigm,chang2012visualization,suzuki2016single}, and Forster resonance energy transfer (FRET)~\cite{de2005lipid} contribute to a deeper understanding of membrane lateral heterogeneity and functions,  however, the nature of lipid rafts and the mechanisms at the origin of their formation are still currently debated~\cite{sezgin2017mystery}. 
The major experimental challenge remains indeed their direct unambiguous microscopic detection \textit{in vivo}, which involves highly complex short-lifetime nanoscale processes.

It has been often speculated that lateral heterogeneities in membranes are caused by the separation of lipids in two coexisting domains with different degree of order (liquid-ordered, with high packing, and liquid-disordered),  lateral diffusivity and composition~\cite{simons2004model,simons2010revitalizing,sezgin2017mystery}. 
This behavior has been observed in artificial membrane model systems, 
from simple bilayers~\cite{tamm1985supported} with few proteins to vesicles~\cite{veatch2003separation,kahya2005raft} up to giant protein-rich membranes~\cite{levental2009cholesterol}.
Phase separated domains are, however, not representative of rafts $\textit{in vivo}$, which occur in a mixed-phase, so that a direct translation of the results obtained in artificial membranes
to living cell membranes remains a major challenge. Model membranes dramatically differs from living membranes, mainly in composition and environmental conditions: Model membranes are at equilibrium and the phase domain partitions strongly depend on the preparation protocol~\cite{de2005lipid,Hancock2006,levental2011raft,sezgin2017mystery}. 

Note that in the present work only flat lipid membranes are discussed and studied. 
Biological membranes, however, are often curved on different length scales~\cite{tian2009sorting} and it has been widely shown 
- experimentally on artificial~\cite{heinrich2010dynamic} and supported membranes~\cite{parthasarathy2006curvature} and theoretically via numerical modeling~\cite{cooke2006coupling,risselada2009curvature} -
that membrane curvature can control the geometry, size and spatial organization of lipid rafts or micro-domains, as discussed e.g. in Ref.~\cite{rozycki2008stable}.

Several theoretical models have been proposed to explain the nanoscale lateral inhomogeneities in flat membranes. A well-known theoretical scenario~\cite{Honerkamp-Smith2008,veatch2008critical,Honerkamp-Smith2009,machta2012critical,Honerkamp-Smith2012,Chen2013,mitra2018computation}, describes the phase transition from liquid-order to liquid-disorder depending on temperature 
in biomimetic membranes, stating that the cell membrane composition is tuned in the proximity of a critical point, where 
fluctuations typical of 2D Ising-like phenomena are observed. 
Other proposed models consider the constitution of microemulsions~\cite{gompper1990lattice}, 
or the presence of nano- to microsized domains in the membrane~\cite{feigenson2001ternary,silvius2003fluorescence,bunge2008characterization}, 
or three-component fluid mixtures that contain an active agent~\cite{Hancock2006}. 
Further approaches examine the role of the underlying cytocortex and the actin cytoskeleton~\cite{Chichili2009,Machta2011,head2014interaction,schneider2017diffusion} in the raft organization. 

To shed light on the mechanism leading to the raft formation $\textit{in vivo}$, we perform extensive computer simulations using molecular dynamics (MD) coarse-grained model proposed by Deserno \textit{et al.}~\cite{Cooke2005} and an Ising-based model \cite{machta2012critical,Chen2013,mitra2018computation}. The advantages of this approach are the 
direct investigation of raft dynamics, in terms of stability and mobility, and the possibility to explore a wide range of time and length scale comparing two different methods.

To reproduce living membrane conditions, we set both models in a mixed phase, in the proximity of the transition point, carefully checking for the absence of already phase-separated domains. We find that the introduction of a transmembrane raft protein, interacting with membrane lipids, triggers the lipids nucleating a local phase separation, in agreement with recent experimental results in living systems~\cite{Komura2016,suzuki2018}. We study the evolution of the raft local domain in terms of average cluster size as a function the variables of the models, like temperature and interaction strength among protein and lipids.  The stability of the average cluster size together with the diffusive motion of the protein observed through the analysis of the mean squared displacement (MSD) indicates that the formation of the raft occurs in the liquid phase.

From real space distributions calculations, we observe that the additional protein drives  the aggregation behavior of the raft. Our analysis, combining different time and length scales - coarse-grained simulations are limited compared to Ising-like approach - suggests 
the fundamental role of transmembrane proteins in the raft formation near the transition point. In analogy with the Ising model, far from the transition line, when the degree of the disorder is too strong, the correlation length is small and therefore the protein is not able induce raft formation. 
Close to the critical point, however, small perturbations as those introduced by the raft protein
can lead to large changes and trigger the formation of a lipid raft.

The paper is organized as follows:
In Sec.~\ref{chap:methods}, we describe our simulation models and their computational setting.
In Sec.~\ref{chap:results}, we show our numerical results.
In Sec.~\ref{chap:conclustion}, we conclude.

\section{Computational methods}\label{chap:methods}
In order to describe the membrane dynamics in a general way, we distinguish between ``raft lipids'' (R), which are those that have the tendency to aggregate in raft and ``non-raft lipids'' (N) that do not assemble in aggregates.
The addition of a transmembrane raft protein (P), interacting with both types of membrane lipids, i.e. R and N, triggers a local phase separation in the form of lipid clusterization.  Both systems are binary mixtures of lipids, R and N, in a $1:7$ concentration ratio. Note that the protein is attractive to the minority phase (e.g. the raft lipids).
The concentration ratio used here is similar to those used in experimental studies on domain emergence in model membranes.  If we wish to compare our models with cellular membranes
considering cholesterol as the main actor in lipid rafts formation,
the ratio 1:7 is too small to describe the real situation in plasma membranes of mammalian 
cells, but might be more relevant for the endoplasmic reticulum or the Golgi apparatus.

\subsection{Molecular dynamics simulations}\label{chap:MD}
For the molecular dynamics model, we simulate a two components bilayer membrane using the solvent-free model introduced by Deserno \textit{et al.}~\cite{Cooke2005}, already employed to reproduce self-assembly, raft formation, melting, and fusion of single and multi component bilayer membranes~\cite{Cooke2005,starr2014dynamical}.  In this model, a lipid is composed of three beads, one representing the hydrophilic head and two the hydrophobic tail. We model the interactions by reduced Lennard-Jones units, where $\sigma$ is the unit of length, $\epsilon$ the unit of energy, $m$ the mass, and time is in units $\tau=\sqrt{\frac{\epsilon}{m\sigma^2}}$~\cite{Andersen1971}. The simulations are performed with the LAMMPS MD simulator~\cite{Plimpton1995} 
and the relevant scripts are available on GitHub~\cite{github_lammps}.
Following previous numerical work~\cite{starr2014dynamical} one can map the units of the model into physical units, taking into account the characteristic scales of lipid bilayers: Considering a membrane thickness of about 5~nm, one has $\sigma \sim$0.5~nm; the diffusion coefficient in the fluid phase is $\sim$1~$\mu$m$^2$s$^{-1}$, then $\tau\sim$10~ns~\cite{Cooke2005,cooke2005solvent}. Considering the transition between ordered and disordered fluid states around 315~K, one can set $\epsilon\sim$4.6~kJ~mol$^{-1}$~\cite{tristram2004lipid,Cooke2005,starr2014dynamical}.

The total interaction potential is given by the sum of four terms:
i)~A potential acting between head-head, head-tail and tail-tail, given by $V(r,b) = 4 \epsilon[(b/r)^{12}- (b/r)^{6} + 1/4] $ with the following set of parameters $b_\mathrm{tail,tail}=\sigma$, $b_\mathrm{head,head}= b_\mathrm{head,tail}=0.95\sigma$. 
These potentials are truncated at a distance $r_\mathrm{c} = 2^{1/6}b$.
ii)~A finite extensible non-linear elastic (FENE) potential, keeping together the three beads: 
$V_\mathrm{bond}(r) =- 1/2 k_\mathrm{bond} r^2_\infty \mathrm{ln}[1-({r}/{r_\infty})^2]$,
where $k_\mathrm{bond} = 30 \epsilon/\sigma^2$ is the  stiffness  and the divergence length $r_\infty = 1.5\sigma$.
iii)~An harmonic spring potential with rest length $4\sigma$ between the head and the lower tail bead: $V_\mathrm{bend}(r)=1/2 k_\mathrm{bend} (r - 4 \sigma)^2$.
iv)~An attractive potential that only affects the tail-tail interaction, given by
\begin{equation}\label{eq:attrPot}
V_\mathrm{attr}(r) =
\begin{cases} 
- \epsilon  \quad & \left( r < r_\mathrm{c} \right)\\ 
- \epsilon  \cos ^2 \left( \frac{\pi \left( r - r_\mathrm{c} \right)}{ 2 w_\mathrm{c}} \right) \quad & \left( r_\mathrm{c} \leq r \leq r_\mathrm{c} + w_\mathrm{c} \right) \\
0 \quad & \left( r > r_\mathrm{c} + w_\mathrm{c} \right) 
\end{cases}
\end{equation}
where $w_\mathrm{c}$ is the potential width, the key tunable parameter, together with the temperature, determining the phase-state of the lipid species in the membrane~\cite{Cooke2005}. The tail-tail potential has an attractive regime depending on the potential width $w_\mathrm{c}$.

We perform equilibration runs and prepare the system with the two types of lipids in a mixed fluid phase by tuning the potential widths $w_\mathrm{c}^\mathrm{RR}$=$w_\mathrm{c}^\mathrm{NN}$=1.8$\sigma$ and $w_\mathrm{c}^\mathrm{RN}$=1.75$\sigma$. Note that increasing the difference between the potential widths $w_\mathrm{c}^\mathrm{RR,NN}$ and $w_\mathrm{c}^\mathrm{RN}$ leads to phase-separation of the two types of lipids, not suitable to represent physiological membrane conditions.  We simulate a double layer with 4100 lipids in a box size with periodic boundary conditions along $x$ and $y$ in a NpT ensemble, controlled by a Nos\'e-Hoover thermostat. The target pressure is zero. 
We investigate four different temperatures in the liquid phase, from 1.10 to 1.25 (reduced units). Note that the range of temperature is limited in order to preserve the stability of the membrane and the mixed phase condition.

\begin{figure}
	\centering
	\includegraphics[scale=1.0]{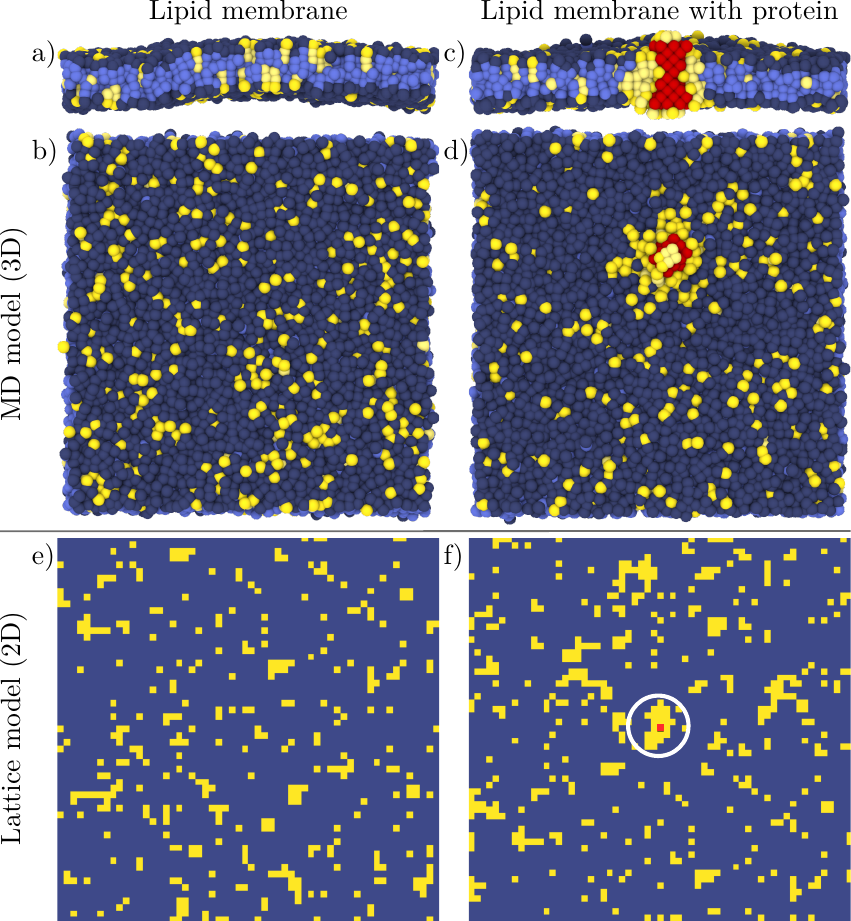}
	\caption{
		\label{fig:figure2}
		\textbf{Protein-driven formation of lipid raft domain in a binary mixture membrane in a mixed phase.}
		a)~Lateral and b)~top view of the bilayer in a mixed phase simulated with coarse-grained MD model. Yellow beads represent raft lipids, blue beads non-raft lipids. Lipid heads are indicated with darker colors. 
		c)~Side view of the protein (red) in MD simulations. 
		d)~Top view of the lipid raft (cluster of yellow lipids) surrounding the protein (red) in MD simulations. 
        e)~Top view of the lipid raft (cluster of yellow lipids) from the 2D Ising-like simulations, blue dots correspond to N lipids.
        f)~Top view of the lipid raft in 2D Ising-like simulations with the insertion of the protein (red dot).
        See videos of {MD model}~\cite{S1_MD_video} and {lattice model}~\cite{S2_lattice_video}  simulations showing a flat membrane in the mixed phase where a protein induces the formation of clusters.
	}
\end{figure}

The lateral view of the bilayer structure, reported in Fig.~\ref{fig:figure2}(a), shows raft lipids with yellow beads and non-raft lipids with blue beads. The heads (marked with dark yellow and dark blue) point outside the leaflet while the tails face towards the center. 
The top view of the membrane in the mixed phase at $T$=1.10 is presented in Fig.~\ref{fig:figure2}(b).

After equilibration runs, we introduce an additional transmembrane protein that is attractive to raft lipids and lead to their local clusterization.  The protein interacts with both types of lipids with a simple Lennard-Jones potential, as done elsewhere~\cite{klingelhoefer2009peptide}:
$V_\mathrm{LJ} \left( r \right) = 4 \bar{\epsilon} [( \frac{1.2\sigma}{r} )^{12} -(\frac{1.2\sigma}{r})^{6}] $ for $r\leq r_\mathrm{c}$. 
The depth of the potential well is $\bar{\epsilon}=2$ and the cutoff distance $r_\mathrm{c}=6.0\sigma$ for the interaction with lipids of type R, while $\bar{\epsilon}=0.05$ and $r_\mathrm{c}=2.0\sigma$ for the interaction with lipids of type N.
We model the protein with 78~beads arranged in a face-centered cubic structure with a hourglass shape, so that the two extremes have a radius of $\sim 2.5\sigma$
that linearly reduces to $\sim1.0\sigma$ at the center.
The lateral extension of the protein ($\sim7\sigma$) is slightly bigger than the thickness of the membrane ($\sim6\sigma$), in agreement with previous works~\cite{klingelhoefer2009peptide}. The lateral and top views of the protein embedded in the membrane, triggering a raft domain are reported in Fig.~\ref{fig:figure2}(c,d) for the system at $T$=1.10. The beads  constituting the protein are represented in red and they are treated as a single rigid body, i.e. they maintain their relative position throughout the MD dynamics. therefore the protein egg timer shape is always conserved. Simulations are performed for a membrane composed by 40000 lipids.
The evolution dynamics of the protein driven lipid raft domain at $T$=1.10 is shown in Ref.~\cite{S1_MD_video}.

\subsection{Lattice model}\label{chap:Ising}
To mimic a two dimensional cell membrane, we also use an Ising-based lattice model. 
The advantage of this approach, compared to MD simulations studied in this paper, is the very low computational cost that enable us to explore a wide phase diagram on longer timescales.
The code of the model for all simulations preformed is available at GitHub~\cite{github_lattice_model}.
The membrane of the cell is represented by a two-dimensional square lattice of size $N \times N$ with spacing set to $1$. Simulations are performed on a lattice of size $N=128$.
Periodic boundary conditions are applied in both dimensions. Each lattice site $c_i$ is occupied by either a lipid of type R or N, in analogy with the spins in the Ising model that can have either state ``up'' or ``down''.
Like in the Ising model, each lipid has pairwise interaction within a distance of one lattice spacing, therefore it can only interact with the four-nearest-neighboring lipids (left, right, top and bottom).
The total energy of the system is represent by the Hamiltonian:
\begin{equation}\label{eq:hamiltonian}
	H \left( \left\lbrace c_i \right\rbrace_{i=1}^{N^2} \right) =  \sum_{\left\langle i,j\right\rangle} E_\mathrm{c_i,c_j}
\end{equation}
where the sum is over all nearest-neighbor pairs and $E_\mathrm{c_i,c_j}$ is the energy of the interaction of the lipids on lattice sites $i$ and $j$ of type $c_i$ and $c_j$.
Interaction between two lipids of the same type (RR or NN) is attractive ($E_\mathrm{RR} = E_\mathrm{NN} < 0$), whereas the interaction among different type of lipids (RN or NR) is repulsive ($E_\mathrm{RN} >0$).
The interaction energies are related by $E_\mathrm{RR} = E_\mathrm{NN} = - E_\mathrm{RN}$.
The interaction strength is always given as absolute values of the lipid-lipid interaction energies $\ERN$ and in units of $\frac{1}{\beta}$ where $\beta = \left( k_\mathrm{B} T\right)^{-1}$ .

We simulate the model using Monte Carlo simulations with conserved dynamics~\cite{krapivsky2010kinetic,Kawasaki1965} so that
each lipid can exchange position with one of its four nearest-neighbors. The exchange of two lipids 
from state $\left\lbrace c_i \right\rbrace$ to state  $\left\lbrace c_i^\prime \right\rbrace$ follows the Glauber dynamics~\cite{Glauber1963} and is executed with probability
\begin{equation}\label{eq:glauberDyn}
	p \left( \Delta H \right) = \frac{1}{1 + e^{\beta \Delta H}}
\end{equation}
where $\Delta H = H\left( \left\lbrace c_i^\prime \right\rbrace \right) -  H\left( \left\lbrace c_i  \right\rbrace \right)$ ~\cite{Glauber1963}. At equilibrium, the critical interaction strength 
for the Ising model is $\ERN = \frac{\ln(1 + \sqrt{2})}{2}\approx 0.44$ ~\cite{Baxter_1982}.
Below the critical interaction strength, the lipids are in a mixed phase and above they phase separate into two domains. 
Here we focus on the mixed phase and therefore we limit the range of the interaction strength to
\begin{equation}\label{eq:condRN}
	\ERN \leq 0.4.
\end{equation}
A snapshot of the membrane in the mixed phase is reported in Fig.~\ref{fig:figure2}(e). In the following,
a single Monte Carlo step corresponds to a sequence of $N^2$ updates of randomly selected nearest neighbor pairs 
\cite{Honerkamp-Smith2012,Plischke2006,Metropolis1953}. 

To induce raft formation we add a protein in the membrane by replacing one lipid R. 
The protein interacts with both R and N lipids, therefore we introduce the nearest-neighbor interaction energies $E_\mathrm{RP}$ and $E_\mathrm{NP}$.
To induce a local aggregation of raft lipids, the protein is attractive to lipids of type R ($E_\mathrm{RP} < 0$) and it is repulsive to lipids of type N ($E_\mathrm{NP} > 0$). For simplicity, we chose the two interaction energies as $E_\mathrm{RP} = - E_\mathrm{NP}$ and impose 
\begin{equation}\label{eq:condRPRN}
	\ERP \geq \ERN,
\end{equation}
which promotes clustering of $R$ lipids. The protein evolves according to the same dynamics of the lipids.
A snapshot of the protein-driven raft formation in the lattice model is reported in Fig.~\ref{fig:figure2}(f): The protein, represented with a red square, is surrounded by a cluster of raft lipids (yellow) as highlighted by the white circle. The evolution dynamics of the protein driven lipid raft domain in the lattice model is shown in Ref.~\cite{S2_lattice_video}.
\section{Results}\label{chap:results}
To clarify the effect of the presence of the protein on the raft formation, we investigate for both models the following quantities:
i) The size of aggregated raft lipid clusters around the protein, ii) the dynamical behavior of the protein by MSD, and iii) 
the real-space organization of the raft.

\subsection{Cluster size}\label{chap:clustersize}
\begin{figure}
	\centering
	\includegraphics[scale=1.]{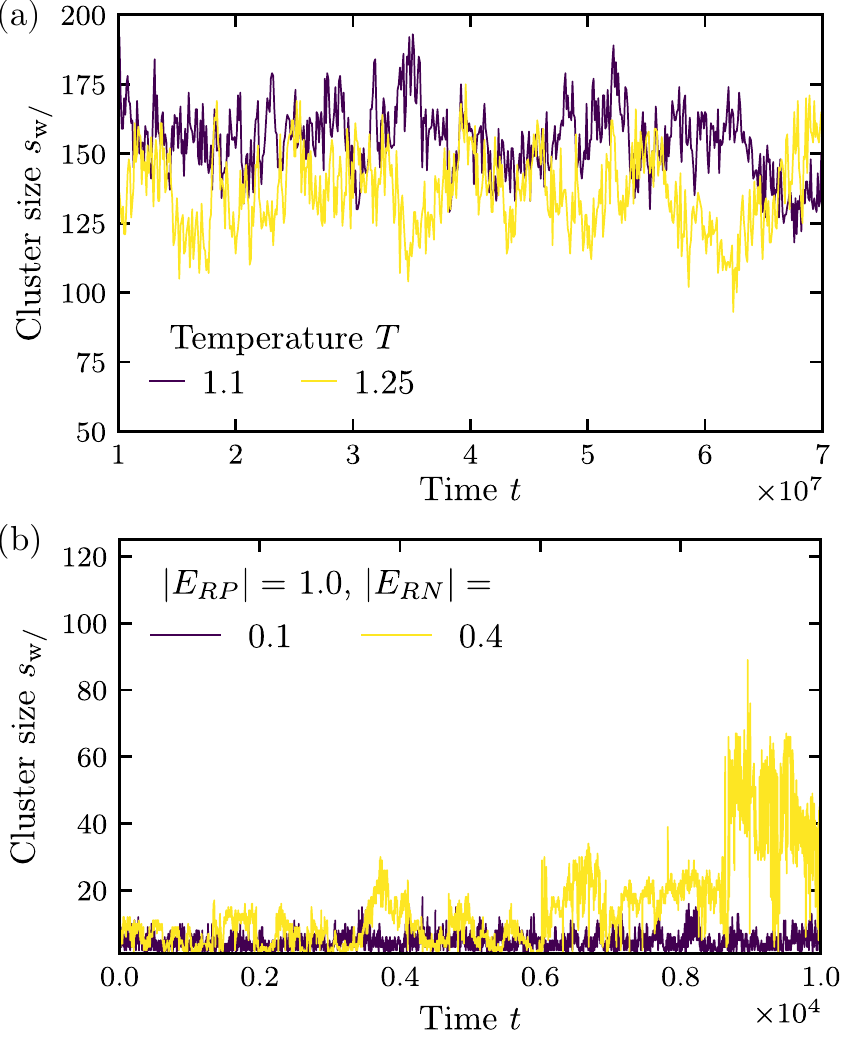}
	\caption{
		\label{fig:figure3_1}
		\textbf{Evolution of the cluster size of lipid rafts with time}.
		Both panels show the evolution of the cluster size of lipid rafts with ( ``w/ protein'') a protein versus time depending on the temperature $k_\mathrm{B}T/\epsilon$ and interaction strength $\ERN$, respectively, for the MD (a) ~and lattice (b) model.
	}
\end{figure}
\begin{figure}
	\centering
	\includegraphics[scale=1.]{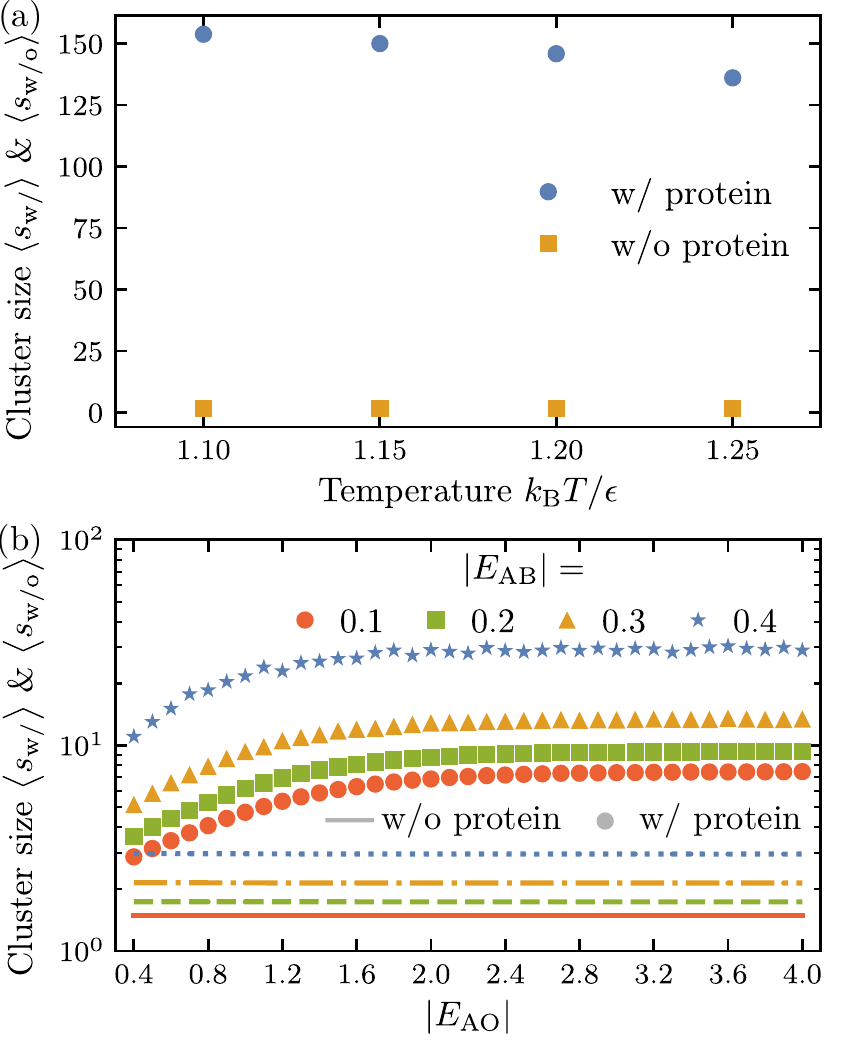}
	\caption{
		\label{fig:figure4}
		\textbf{Average cluster size of lipid rafts}.
		Both panels show the average cluster size of lipid rafts with ( ``w/ protein'') and without ( ``w/o protein'') a protein depending on the temperature $k_\mathrm{B}T/\epsilon$ and interaction strength $\ERP$, respectively, for the MD (a) ~and lattice (b) model.
	}
\end{figure}
In the MD model, a cluster is defined as a set of connected raft lipids that fulfill the neighboring criterion,  based on the mutual distance among all beads that form the lipids. In particular, two raft lipids belong to the same
cluster if their distance is less than $1.2\sigma$.
After the addition of the transmembrane protein, raft lipids aggregate around it until the cluster size reaches a steady state.

Fig.~\ref{fig:figure3_1}(a) shows the evolution of the cluster size with time as a function of two rescaled temperature~$k_\mathrm{B}T/\epsilon$ for the MD model, while Fig.~\ref{fig:figure3_1}(b) the variation in time for different interaction strength $\ERN$,
The cluster size of the cluster containing the one obstacle $s_{w/}$ fluctuates strongly in time in the lattice model with respect to the MD simulations. The time needed for the cluster to grow in the lattice model, gets longer with increasing $\ERP$ and $\ERN$. Specifically we find that increasing~$\ERN$, the persistence time of the cluster grows.

Fig.~\ref{fig:figure4}(a) shows the time-averaged cluster size as a function of the rescaled temperature~$k_\mathrm{B}T/\epsilon$:
the cluster of lipids around the protein, indicated with the blue dot markers (labeled as ``w/ protein''), decreases with increasing temperatures. 
Raft lipids also clusterize without the assistance of the protein, but the
cluster size is a factor of ten smaller, as indicated by the yellow squares (labeled as ``w/o protein''). 

For the lattice model, two raft lipids are assigned to the same cluster if they are nearest neighbors on the lattice. Clusters
are identified using the Hoshen-Kopelmann algorithm for percolation~\cite{Hoshen1976}.
In this model description, the size of clusters strongly fluctuates in time, and clusters are subjected to continuous formation and disruption,  in analogy with experimental observation of fluctuating raft formation, as reported in recent experiments~\cite{Komura2016,suzuki2018}.

In Fig. \ref{fig:figure4}(b) after the addition of the protein, we observe that the clusters of lipids around the protein, indicated with markers (``w/ protein''), saturate with increasing interaction strength between the lipids and the protein $\ERP$: The size of clusters involving a protein increases up to maximum size at $\ERP\approx 3$ and remains constant for stronger interaction strengths. With increasing interaction strengths between the lipids $\ERN$, the average cluster size increases.
For completeness, we also report the cluster size behavior of lipids without the presence of the protein, indicated indicated with lines~(``w/o protein''). 

Since the effective temperature $k_\mathrm{B}T/\epsilon$ scales as the inverse of the interaction strength $\ERN$, the average cluster size decreases in both the MD model and lattice model with increasing temperature, coherently confirming the trend on two different lengthscales and timescales.

\subsection{Diffusion}\label{chap:diffusion}
\begin{figure}
	\centering
	\includegraphics[scale=1.]{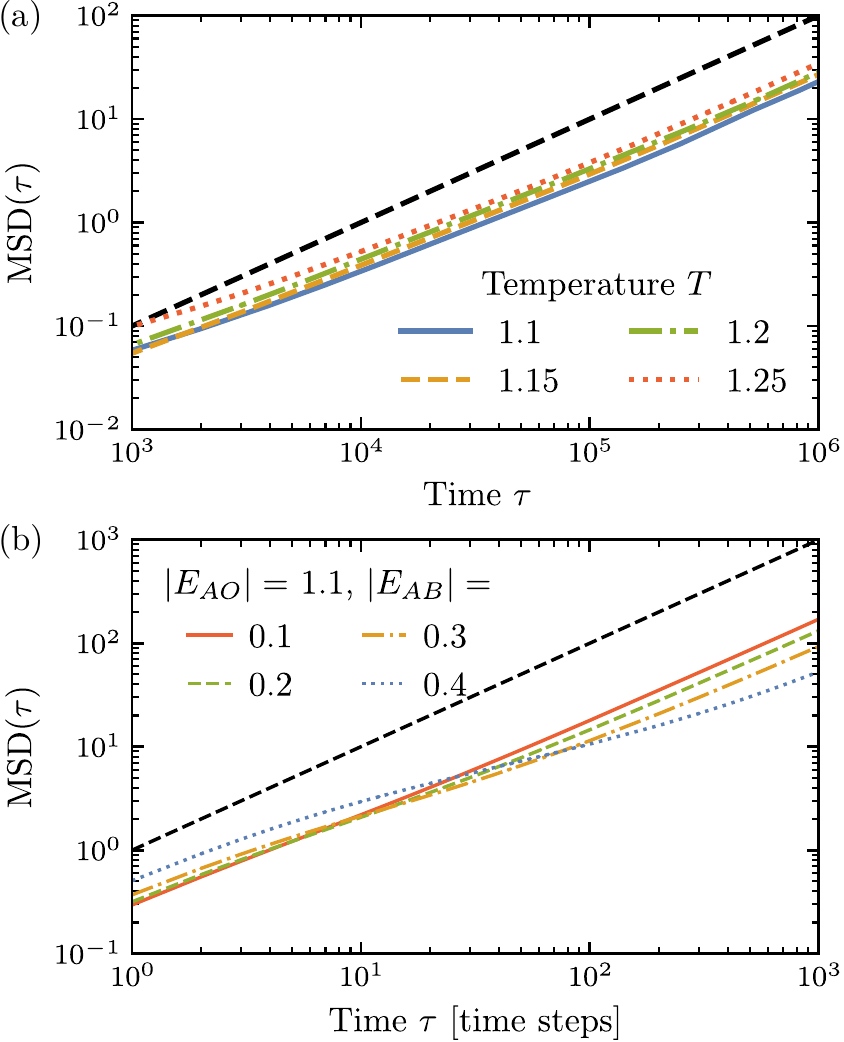}
	\caption{
		\label{fig:figure5}
		\textbf{Mean Squared Displacement of the protein} 
		(a)~In MD simulations, the protein diffuses in the liquid phase.
		(b)~In the lattice model, at short timescales the protein diffuses in the cluster whereas at long timescales it diffuses with the whole cluster.
	}
\end{figure}

The analysis of the mean squared displacement reveals information on the type of diffusion regimes and enable the determination of the diffusion constant. In this section we report the MSD of the protein to examine its mobility with the aggregated cluster.
For both models, we focus on the protein movement within the two-dimensional surface of the lipid membrane. The MSD is defined as
\begin{equation}\label{eq:defMSD2D}
\mathrm{MSD}\left( \tau \right) = 
\left\langle \lvert \vec{x}\left( t + \tau \right) - \vec{x} \left( t \right) \rvert^2 \right\rangle_{t}
\end{equation}
depending on the time increment $\tau$ where $\vec{x}\left( t \right)$ is the position in the membrane and $\langle ~ \cdot ~ \rangle_t$  is the average over  time $t$. In order to calculate the MSD in (\ref{eq:defMSD2D}), $t$ is chosen such that the membrane is at equilibrium, i.e. he average cluster size is stable. Fig.~\ref{fig:figure5} shows the MSD as a function of time for the (a)~MD - and (b)~lattice model.

In the MD simulations, the position of the protein is the center of mass in the $xy$-plane. Fig.~\ref{fig:figure5}(a) shows that the behavior of MSD changes over time  at various temperatures. Although the P mobility is suppressed with decreasing temperature, the overall motion is diffusive, confirming that the state point is in the liquid phase.

In order to determine the type of diffusion and the corresponding diffusion coefficient, we fit the data with the following function
\begin{equation}
\mathrm{MSD} \left( \tau \right) = 4 D \tau^{2H}
\end{equation}
where $H$ is the Hurst exponent, which characterizes the type of diffusion:
$H=\frac{1}{2}$ is linear, $H<\frac{1}{2}$ sub-diffusive and $H>\frac{1}{2}$ super-diffusive behavior \cite{Hurst1951,Mandelbrot1968}. Notice that D corresponds to the usual diffusion constant only for Brownian diffusion \cite{Metzler2000,Taloni2010}.
For the four investigated temperatures where the membrane is in the liquid phase, we find that a fit in the interval $\tau$=10$^3$-10$^4$ yields a subdiffusive regime as   H$\approx$0.4. For $\tau$=10$^5$-10$^6$,  H$\approx$0.5 pointing out that in the long run the diffusion is normal. As a matter of fact, the black dashed line is a reference line for the MSD curve expected for a linear diffusion process with diffusion constant $D =2.5 \cdot 10^{-5}$.

In the lattice model, after a transient regime the protein displays a diffusive behavior, in analogy of what has been observed in MD simulations. The crossover is ultimately ascribable to the different dynamical regimes followed by P: For short times the protein diffuses within the cluster, while on longer timescales the protein diffusion is that of the aggregated cluster. Attempting a linear fit of the asymptotic regime through the formula
\begin{equation}
\mathrm{MSD} \left( \tau \right) = 4 D \tau,
\end{equation}
allows a systematic investigation of the D dependence on the simulations parameters. Indeed, 
we report the MSD behavior with the interaction strength $\ERP$ set to $1.1$ and $\ERN$ varied
in Fig.~\ref{fig:figure5}(b). 
The black dashed line illustrates the MSD of a protein that does not interact with the lipids.
Here the interaction strength among the lipids $\ERN$ is also set to zero.
As the cluster size grows with increasing interaction strength $\ERN$ (see~\ref{fig:figure4}b)), the diffusion constant decreases. 

\subsubsection{Components of the Diffusion Constant}\label{chap:CompDiff}
\begin{figure}
	\centering
	\includegraphics[scale=1.]{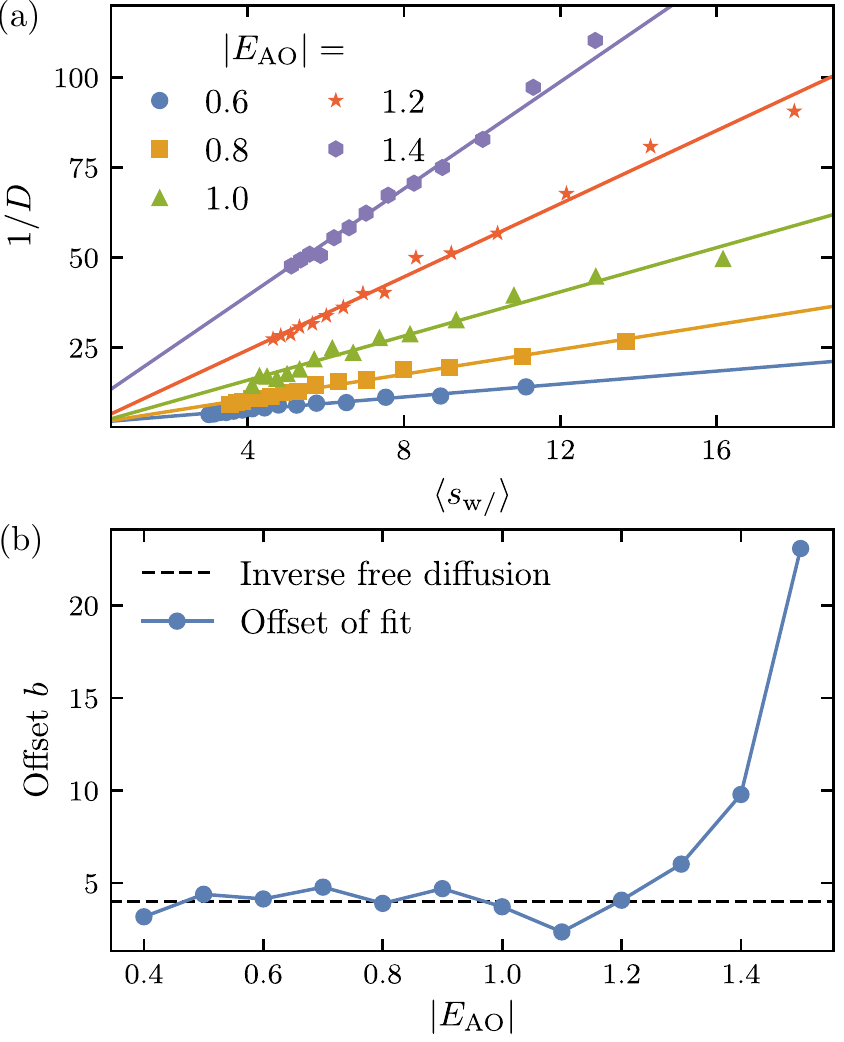}
	\caption{
		\label{fig:figure6}
		\textbf{Diffusion constant analysis for the lattice model.}
		a)~Inverse diffusion constant as a function of the cluster size for fixed interaction strengths $\ERP$ showing a linear behavior (lines represent linear fits).
		The fitting function is $1/D_\mathrm{fit}\left( \left\langle s_\mathrm{w/} \right\rangle \right) = a \cdot \left\langle s_\mathrm{w/} \right\rangle + b$.
		b)~Offset~$b$ of the fitted curves presented in a) as a function of $\ERP$.
		For $\ERP < 1.5$, the offset is quite constant around 4 which corresponds to the inverse of the diffusion constant of a non-interacting protein ($\ERN = 0$ and $\ERP = 0$).
	}
\end{figure}

In the conventional description of Brownian motion, the diffusion constant is proportional to the inverse  mass which here is quantified by the cluster size. We therefore assume $D \sim \left\langle s_\mathrm{w/} \right\rangle^{-1}$, 
and introduce the fitting function
\begin{equation}
D \left( \left \langle s_\mathrm{w/} \right \rangle \right) =
\frac{1}{a \cdot \left \langle s_\mathrm{w/} \right \rangle + b}
\end{equation}
where $a$ is the angular coefficient and $b$ the offset. 
For the sake of clarity, Fig.~\ref{fig:figure6}(a) shows the inverse diffusion constant as a function of the average cluster size. For different values of the interaction strength $\ERP \in \left\{ 0.6, 0.8, 1.0, 1.2, 1.4 \right\}$, indicated by different markers which indicate the simulation data and the solid line is the fit.  The fit works quite well for all interaction strengths $\ERP < 1.5$. Fig. \ref{fig:figure6}(b) shows the offset $b$ as function of the interaction strength $\ERP$.
For $\ERP < 1.4$, the offset is quite constant around $4$.

The offset means that the diffusion also has a constant part, that is  not influenced by the cluster size of the cluster with the protein. For weaker interaction strengths $\ERP$, the offset becomes important since the cluster size decreases, as well as the slope $a$.
So if the interaction strength go to $\ERP \rightarrow 0$ and $s \rightarrow 0$, only the constant is left and the inverse diffusion constant become $\frac{1}{D} = 4$. By inverting this, we get that the diffusion constant is $D = \frac{1}{4}$ that is the diffusion of a non-functional protein.

Hence, we conclude that the diffusion process has two contributions:
One that is constant and is related to the free diffusion of the protein in the cluster, 
while the second depends on the cluster size and is due to the cluster diffusing in the system. 

\subsection{Real space distribution}\label{chap:distribution}
Real space distribution functions give us information about the variation in density relative to a reference point which is the protein in our models. 
For that, we focus on the radial distribution function (RDF) which indicates the density of raft lipids depending on the distance to  the protein.
In general, we normalize the RDF by homogeneous distribution of raft lipids according to the respective lipid ratios, so that values greater (smaller) than $1$ indicate  higher (lower) density than an equilibrated system.
\begin{figure} % real space distribution
	\centering
	\includegraphics[scale=1.]{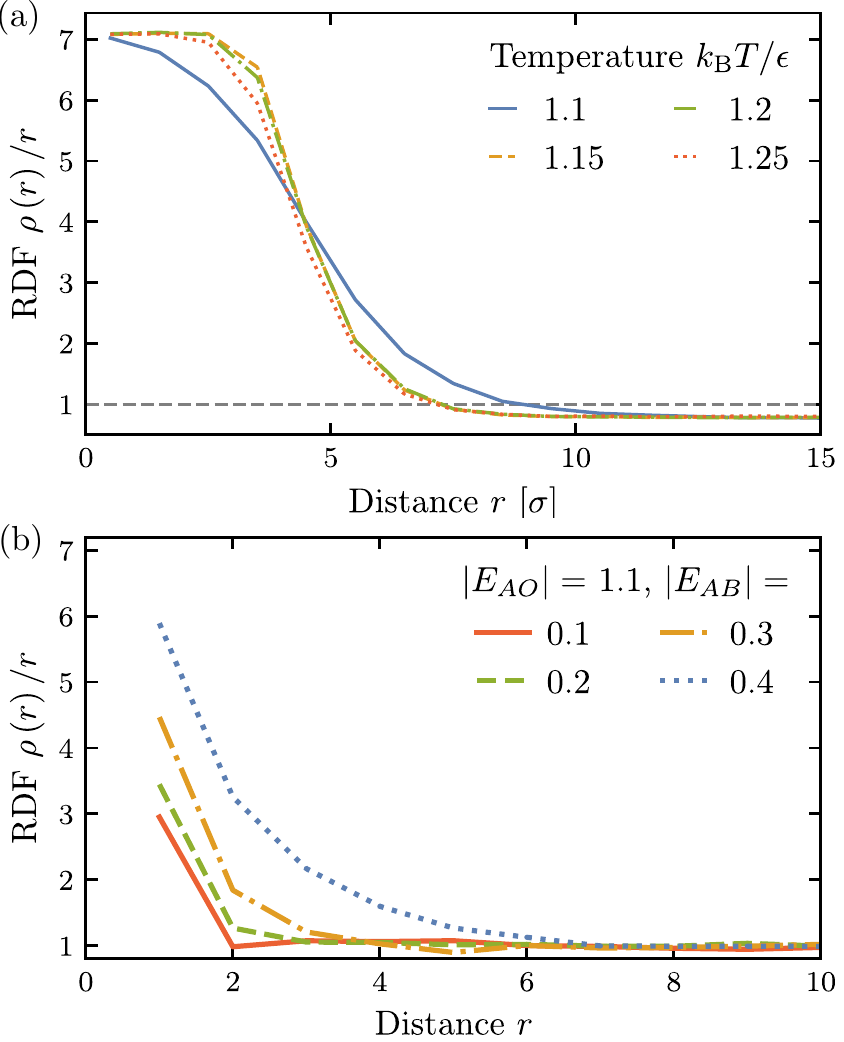}
	\caption{
		\label{fig:figure7}
		\textbf{Time-averaged RDF of lipids around the protein}
		(a)~MD simulations and (b)~Lattice model.
		With decreasing the a)~temperature or b)~increasing the attraction~$\ERN$, the tail of RDF is growing, reflecting that the size of the raft is increasing.
	}
\end{figure}
Fig.~\ref{fig:figure7}(a) shows the RDF of raft lipids relative to the center of mass of the protein in the $xy$-plane for the MD model.
Considering the membrane to be equilibrated with a stable cluster, we report the time-average RDF.
We note that when decreasing the temperature the tail of the RDF increases, consistently with the increase of the average cluster size, observed in Fig.~\ref{fig:figure4}(a). 

Fig.\ref{fig:figure7}(b) shows the RDF of raft lipids around the protein for a fixed interaction strength $\ERP=1.1$. The distribution function shows a high density close to the protein which decays toward the value 1. With increasing interaction strength $\ERN$, the maximum density close to the protein increases in agreement with the results shown in Fig.\ref{fig:figure4}(b). 

\section{Conclusions}\label{chap:conclustion}
Lipid rafts are microdomains on the plasma membrane enriched in cholesterol, sphingolipids and GPI-anchored proteins who play a crucial  role in many cellular processes, such as signal transduction, membrane trafficking, and pathogen entry. Moreover, the aggregation of these microdomains allows to display a cellular response in a very short time. 
The mechanisms underlying rapid raft formation of have been long debated \cite{sezgin2017mystery}.

In this paper, we have studied the role of a transmembrane protein on lipid raft  formation by means of coarse-grained molecular dynamics simulations and an Ising model-based approach. Our results show that the protein triggers the lipid aggregation in the form of a raft in agreement with experiments \cite{Komura2016,suzuki2018}. Raft formation in our model occurs in the disordered phase and does not reflect stable phase separation between raft lipids. From the point of view of statistical mechanics of phase transitions, lipid rafts could be seen as droplets of the raft lipids in the mixed phase. In  the Ising model, those droplets would be the size of the correlation length $\xi$ which becomes large close to the critical point, but would be very short-lived. The action of the transmembrane protein is analogous to a localized magnetic field in the Ising model which help nucleate and stabilize the droplet. 

Our model is thus in agreement with experimental observations in living cells suggesting that lipid raft are dynamic structures rather than stable domains as observed in synthetic lipid membranes \cite{sezgin2017mystery}.  Our results also show that the size of the cluster of raft lipids surrounding the protein reaches a steady state and decreases: i)~With increasing the temperature in the MD model ii)~with the inverse of the interaction strength  (which we can consider as a proxy for temperature) in the lattice simulations.  
Note that our results do not strictly depend on the concentration of the phases as long as the parameters of the system place it at the boundary of the mixed phase. The cluster sizes will only be rescaled by a concentration dependent factor, but the general phenomenology we describe should remain unchanged.

Furthermore, from MSD analysis we confirm that the overall motion of the protein is diffusive in both models. The diffusion constant depends on the size of the raft domain, indicating that the raft and and the protein move coherently as a single object. The results obtained combining two complementing simulation methods tackling different length and time scales coherently support a protein-driven lipid raft formation model.  

Our model treats interactions between lipids in a generic way and could thus be applied to lipid rafts but also to other cases. For instance, evidences of aggregation of lipids around a protein forming a ring or annular shell have been reported by electron spin resonance (ESR) studies since the late seventies~\cite{lee1977annular,watts1979rhodopsin,watts1981distinct,davoust1982simulation,sternberg1992essential}.
and might also be described by our model.

\section{Acknowledgments}
We thank F. Brauns, C.-F. Chou, E. Frey, R. Guerra, C. Negri, A. L. Sellerio for useful discussions.
MH was supported by the ERASMUS+ program under a joint agreement between the University of Milan and Ludwig Maximilian University Munich.
Authors contributions: MH and SB performed numerical simulations. MH,SB,AT analyzed data. CAMLP and SZ designed and  coordinated the project. 

\bibliographystyle{apsrev4-1}
%\bibliography{references_v22}

\begin{thebibliography}{84}%
	\makeatletter
	\providecommand \@ifxundefined [1]{%
		\@ifx{#1\undefined}
	}%
	\providecommand \@ifnum [1]{%
		\ifnum #1\expandafter \@firstoftwo
		\else \expandafter \@secondoftwo
		\fi
	}%
	\providecommand \@ifx [1]{%
		\ifx #1\expandafter \@firstoftwo
		\else \expandafter \@secondoftwo
		\fi
	}%
	\providecommand \natexlab [1]{#1}%
	\providecommand \enquote  [1]{``#1''}%
	\providecommand \bibnamefont  [1]{#1}%
	\providecommand \bibfnamefont [1]{#1}%
	\providecommand \citenamefont [1]{#1}%
	\providecommand \href@noop [0]{\@secondoftwo}%
	\providecommand \href [0]{\begingroup \@sanitize@url \@href}%
	\providecommand \@href[1]{\@@startlink{#1}\@@href}%
	\providecommand \@@href[1]{\endgroup#1\@@endlink}%
	\providecommand \@sanitize@url [0]{\catcode `\\12\catcode `\$12\catcode
		`\&12\catcode `\#12\catcode `\^12\catcode `\_12\catcode `\%12\relax}%
	\providecommand \@@startlink[1]{}%
	\providecommand \@@endlink[0]{}%
	\providecommand \url  [0]{\begingroup\@sanitize@url \@url }%
	\providecommand \@url [1]{\endgroup\@href {#1}{\urlprefix }}%
	\providecommand \urlprefix  [0]{URL }%
	\providecommand \Eprint [0]{\href }%
	\providecommand \doibase [0]{http://dx.doi.org/}%
	\providecommand \selectlanguage [0]{\@gobble}%
	\providecommand \bibinfo  [0]{\@secondoftwo}%
	\providecommand \bibfield  [0]{\@secondoftwo}%
	\providecommand \translation [1]{[#1]}%
	\providecommand \BibitemOpen [0]{}%
	\providecommand \bibitemStop [0]{}%
	\providecommand \bibitemNoStop [0]{.\EOS\space}%
	\providecommand \EOS [0]{\spacefactor3000\relax}%
	\providecommand \BibitemShut  [1]{\csname bibitem#1\endcsname}%
	\let\auto@bib@innerbib\@empty
	%</preamble>
	\bibitem [{\citenamefont {Andersen}\ and\ \citenamefont
		{Koeppe}(2007)}]{andersen2007bilayer}%
	\BibitemOpen
	\bibfield  {author} {\bibinfo {author} {\bibfnamefont {O.~S.}\ \bibnamefont
			{Andersen}}\ and\ \bibinfo {author} {\bibfnamefont {R.~E.}\ \bibnamefont
			{Koeppe}},\ }\href@noop {} {\bibfield  {journal} {\bibinfo  {journal} {Annu.
				Rev. Biophys. Biomol. Struct.}\ }\textbf {\bibinfo {volume} {36}},\ \bibinfo
		{pages} {107} (\bibinfo {year} {2007})}\BibitemShut {NoStop}%
	\bibitem [{\citenamefont {Ramakrishnan}\ \emph {et~al.}(2014)\citenamefont
		{Ramakrishnan}, \citenamefont {{Sunil Kumar}},\ and\ \citenamefont
		{Radhakrishnan}}]{Ramakrishnan2014}%
	\BibitemOpen
	\bibfield  {author} {\bibinfo {author} {\bibfnamefont {N.}~\bibnamefont
			{Ramakrishnan}}, \bibinfo {author} {\bibfnamefont {P.~B.}\ \bibnamefont
			{{Sunil Kumar}}}, \ and\ \bibinfo {author} {\bibfnamefont {R.}~\bibnamefont
			{Radhakrishnan}},\ }\href {\doibase 10.1016/j.physrep.2014.05.001} {\bibfield
		{journal} {\bibinfo  {journal} {Physics Reports}\ }\textbf {\bibinfo
			{volume} {543}},\ \bibinfo {pages} {1} (\bibinfo {year} {2014})}\BibitemShut
	{NoStop}%
	\bibitem [{\citenamefont {Sezgin}\ \emph {et~al.}(2017)\citenamefont {Sezgin},
		\citenamefont {Levental}, \citenamefont {Mayor},\ and\ \citenamefont
		{Eggeling}}]{sezgin2017mystery}%
	\BibitemOpen
	\bibfield  {author} {\bibinfo {author} {\bibfnamefont {E.}~\bibnamefont
			{Sezgin}}, \bibinfo {author} {\bibfnamefont {I.}~\bibnamefont {Levental}},
		\bibinfo {author} {\bibfnamefont {S.}~\bibnamefont {Mayor}}, \ and\ \bibinfo
		{author} {\bibfnamefont {C.}~\bibnamefont {Eggeling}},\ }\href@noop {}
	{\bibfield  {journal} {\bibinfo  {journal} {Nature Reviews Molecular Cell
				Biology}\ }\textbf {\bibinfo {volume} {18}},\ \bibinfo {pages} {361}
		(\bibinfo {year} {2017})}\BibitemShut {NoStop}%
	\bibitem [{\citenamefont {Brown}\ and\ \citenamefont
		{London}(1998)}]{brown1998functions}%
	\BibitemOpen
	\bibfield  {author} {\bibinfo {author} {\bibfnamefont {D.}~\bibnamefont
			{Brown}}\ and\ \bibinfo {author} {\bibfnamefont {E.}~\bibnamefont {London}},\
	}\href@noop {} {\bibfield  {journal} {\bibinfo  {journal} {Annual review of
				cell and developmental biology}\ }\textbf {\bibinfo {volume} {14}},\ \bibinfo
		{pages} {111} (\bibinfo {year} {1998})}\BibitemShut {NoStop}%
	\bibitem [{\citenamefont {Anderson}\ and\ \citenamefont
		{Jacobson}(2002)}]{anderson2002role}%
	\BibitemOpen
	\bibfield  {author} {\bibinfo {author} {\bibfnamefont {R.~G.}\ \bibnamefont
			{Anderson}}\ and\ \bibinfo {author} {\bibfnamefont {K.}~\bibnamefont
			{Jacobson}},\ }\href@noop {} {\bibfield  {journal} {\bibinfo  {journal}
			{Science}\ }\textbf {\bibinfo {volume} {296}},\ \bibinfo {pages} {1821}
		(\bibinfo {year} {2002})}\BibitemShut {NoStop}%
	\bibitem [{\citenamefont {Simons}\ and\ \citenamefont
		{Ikonen}(1997)}]{simons1997functional}%
	\BibitemOpen
	\bibfield  {author} {\bibinfo {author} {\bibfnamefont {K.}~\bibnamefont
			{Simons}}\ and\ \bibinfo {author} {\bibfnamefont {E.}~\bibnamefont
			{Ikonen}},\ }\href@noop {} {\bibfield  {journal} {\bibinfo  {journal}
			{Nature}\ }\textbf {\bibinfo {volume} {387}},\ \bibinfo {pages} {569}
		(\bibinfo {year} {1997})}\BibitemShut {NoStop}%
	\bibitem [{\citenamefont {Lingwood}\ and\ \citenamefont
		{Simons}(2010)}]{lingwood2010lipid}%
	\BibitemOpen
	\bibfield  {author} {\bibinfo {author} {\bibfnamefont {D.}~\bibnamefont
			{Lingwood}}\ and\ \bibinfo {author} {\bibfnamefont {K.}~\bibnamefont
			{Simons}},\ }\href@noop {} {\bibfield  {journal} {\bibinfo  {journal}
			{science}\ }\textbf {\bibinfo {volume} {327}},\ \bibinfo {pages} {46}
		(\bibinfo {year} {2010})}\BibitemShut {NoStop}%
	\bibitem [{\citenamefont {Harder}\ and\ \citenamefont
		{Simons}(1997)}]{harder1997caveolae}%
	\BibitemOpen
	\bibfield  {author} {\bibinfo {author} {\bibfnamefont {T.}~\bibnamefont
			{Harder}}\ and\ \bibinfo {author} {\bibfnamefont {K.}~\bibnamefont
			{Simons}},\ }\href@noop {} {\bibfield  {journal} {\bibinfo  {journal}
			{Current opinion in cell biology}\ }\textbf {\bibinfo {volume} {9}},\
		\bibinfo {pages} {534} (\bibinfo {year} {1997})}\BibitemShut {NoStop}%
	\bibitem [{\citenamefont {Varma}\ and\ \citenamefont
		{Mayor}(1998)}]{varma1998gpi}%
	\BibitemOpen
	\bibfield  {author} {\bibinfo {author} {\bibfnamefont {R.}~\bibnamefont
			{Varma}}\ and\ \bibinfo {author} {\bibfnamefont {S.}~\bibnamefont {Mayor}},\
	}\href@noop {} {\bibfield  {journal} {\bibinfo  {journal} {Nature}\ }\textbf
		{\bibinfo {volume} {394}},\ \bibinfo {pages} {798} (\bibinfo {year}
		{1998})}\BibitemShut {NoStop}%
	\bibitem [{\citenamefont {Brown}(1998)}]{brown1998sphingolipid}%
	\BibitemOpen
	\bibfield  {author} {\bibinfo {author} {\bibfnamefont {R.~E.}\ \bibnamefont
			{Brown}},\ }\href@noop {} {\bibfield  {journal} {\bibinfo  {journal} {Journal
				of cell science}\ }\textbf {\bibinfo {volume} {111}},\ \bibinfo {pages} {1}
		(\bibinfo {year} {1998})}\BibitemShut {NoStop}%
	\bibitem [{\citenamefont {Brown}\ and\ \citenamefont
		{London}(2000)}]{brown2000structure}%
	\BibitemOpen
	\bibfield  {author} {\bibinfo {author} {\bibfnamefont {D.~A.}\ \bibnamefont
			{Brown}}\ and\ \bibinfo {author} {\bibfnamefont {E.}~\bibnamefont {London}},\
	}\href@noop {} {\bibfield  {journal} {\bibinfo  {journal} {Journal of
				Biological Chemistry}\ }\textbf {\bibinfo {volume} {275}},\ \bibinfo {pages}
		{17221} (\bibinfo {year} {2000})}\BibitemShut {NoStop}%
	\bibitem [{\citenamefont {Douglass}\ and\ \citenamefont
		{Vale}(2005)}]{douglass2005single}%
	\BibitemOpen
	\bibfield  {author} {\bibinfo {author} {\bibfnamefont {A.~D.}\ \bibnamefont
			{Douglass}}\ and\ \bibinfo {author} {\bibfnamefont {R.~D.}\ \bibnamefont
			{Vale}},\ }\href@noop {} {\bibfield  {journal} {\bibinfo  {journal} {Cell}\
		}\textbf {\bibinfo {volume} {121}},\ \bibinfo {pages} {937} (\bibinfo {year}
		{2005})}\BibitemShut {NoStop}%
	\bibitem [{\citenamefont {Pike}(2006)}]{pike2006rafts}%
	\BibitemOpen
	\bibfield  {author} {\bibinfo {author} {\bibfnamefont {L.~J.}\ \bibnamefont
			{Pike}},\ }\href@noop {} {\bibfield  {journal} {\bibinfo  {journal} {Journal
				of lipid research}\ }\textbf {\bibinfo {volume} {47}},\ \bibinfo {pages}
		{1597} (\bibinfo {year} {2006})}\BibitemShut {NoStop}%
	\bibitem [{\citenamefont {Owen}\ \emph {et~al.}(2012)\citenamefont {Owen},
		\citenamefont {Williamson}, \citenamefont {Magenau},\ and\ \citenamefont
		{Gaus}}]{owen2012sub}%
	\BibitemOpen
	\bibfield  {author} {\bibinfo {author} {\bibfnamefont {D.~M.}\ \bibnamefont
			{Owen}}, \bibinfo {author} {\bibfnamefont {D.~J.}\ \bibnamefont
			{Williamson}}, \bibinfo {author} {\bibfnamefont {A.}~\bibnamefont {Magenau}},
		\ and\ \bibinfo {author} {\bibfnamefont {K.}~\bibnamefont {Gaus}},\
	}\href@noop {} {\bibfield  {journal} {\bibinfo  {journal} {Nature
				communications}\ }\textbf {\bibinfo {volume} {3}},\ \bibinfo {pages} {1256}
		(\bibinfo {year} {2012})}\BibitemShut {NoStop}%
	\bibitem [{\citenamefont {Schulz}\ \emph {et~al.}(2012)\citenamefont {Schulz},
		\citenamefont {Olubummo},\ and\ \citenamefont {Binder}}]{schulz2012beyond}%
	\BibitemOpen
	\bibfield  {author} {\bibinfo {author} {\bibfnamefont {M.}~\bibnamefont
			{Schulz}}, \bibinfo {author} {\bibfnamefont {A.}~\bibnamefont {Olubummo}}, \
		and\ \bibinfo {author} {\bibfnamefont {W.~H.}\ \bibnamefont {Binder}},\
	}\href@noop {} {\bibfield  {journal} {\bibinfo  {journal} {Soft Matter}\
		}\textbf {\bibinfo {volume} {8}},\ \bibinfo {pages} {4849} (\bibinfo {year}
		{2012})}\BibitemShut {NoStop}%
	\bibitem [{\citenamefont {Honigmann}\ \emph {et~al.}(2014)\citenamefont
		{Honigmann}, \citenamefont {Mueller}, \citenamefont {Ta}, \citenamefont
		{Schoenle}, \citenamefont {Sezgin}, \citenamefont {Hell},\ and\ \citenamefont
		{Eggeling}}]{honigmann2014scanning}%
	\BibitemOpen
	\bibfield  {author} {\bibinfo {author} {\bibfnamefont {A.}~\bibnamefont
			{Honigmann}}, \bibinfo {author} {\bibfnamefont {V.}~\bibnamefont {Mueller}},
		\bibinfo {author} {\bibfnamefont {H.}~\bibnamefont {Ta}}, \bibinfo {author}
		{\bibfnamefont {A.}~\bibnamefont {Schoenle}}, \bibinfo {author}
		{\bibfnamefont {E.}~\bibnamefont {Sezgin}}, \bibinfo {author} {\bibfnamefont
			{S.~W.}\ \bibnamefont {Hell}}, \ and\ \bibinfo {author} {\bibfnamefont
			{C.}~\bibnamefont {Eggeling}},\ }\href@noop {} {\bibfield  {journal}
		{\bibinfo  {journal} {Nature communications}\ }\textbf {\bibinfo {volume}
			{5}},\ \bibinfo {pages} {5412} (\bibinfo {year} {2014})}\BibitemShut
	{NoStop}%
	\bibitem [{\citenamefont {Lorent}\ \emph {et~al.}(2017)\citenamefont {Lorent},
		\citenamefont {Diaz-Rohrer}, \citenamefont {Lin}, \citenamefont {Spring},
		\citenamefont {Gorfe}, \citenamefont {Levental},\ and\ \citenamefont
		{Levental}}]{lorent2017structural}%
	\BibitemOpen
	\bibfield  {author} {\bibinfo {author} {\bibfnamefont {J.~H.}\ \bibnamefont
			{Lorent}}, \bibinfo {author} {\bibfnamefont {B.}~\bibnamefont {Diaz-Rohrer}},
		\bibinfo {author} {\bibfnamefont {X.}~\bibnamefont {Lin}}, \bibinfo {author}
		{\bibfnamefont {K.}~\bibnamefont {Spring}}, \bibinfo {author} {\bibfnamefont
			{A.~A.}\ \bibnamefont {Gorfe}}, \bibinfo {author} {\bibfnamefont {K.~R.}\
			\bibnamefont {Levental}}, \ and\ \bibinfo {author} {\bibfnamefont
			{I.}~\bibnamefont {Levental}},\ }\href@noop {} {\bibfield  {journal}
		{\bibinfo  {journal} {Nature communications}\ }\textbf {\bibinfo {volume}
			{8}},\ \bibinfo {pages} {1219} (\bibinfo {year} {2017})}\BibitemShut
	{NoStop}%
	\bibitem [{\citenamefont {Sezgin}\ and\ \citenamefont
		{Schwille}(2011)}]{sezgin2011fluorescence}%
	\BibitemOpen
	\bibfield  {author} {\bibinfo {author} {\bibfnamefont {E.}~\bibnamefont
			{Sezgin}}\ and\ \bibinfo {author} {\bibfnamefont {P.}~\bibnamefont
			{Schwille}},\ }\href@noop {} {\bibfield  {journal} {\bibinfo  {journal} {Cold
				Spring Harbor perspectives in biology}\ }\textbf {\bibinfo {volume} {3}},\
		\bibinfo {pages} {a009803} (\bibinfo {year} {2011})}\BibitemShut {NoStop}%
	\bibitem [{\citenamefont {Klymchenko}\ and\ \citenamefont
		{{Kreder}}(2014)}]{Klymchenko2014}%
	\BibitemOpen
	\bibfield  {author} {\bibinfo {author} {\bibfnamefont {A.~S.}\ \bibnamefont
			{Klymchenko}}\ and\ \bibinfo {author} {\bibfnamefont {R.}~\bibnamefont
			{{Kreder}}},\ }\href {\doibase 10.1016/j.chembiol.2013.11.009} {\bibfield
		{journal} {\bibinfo  {journal} {Chemistry {\&} Biology}\ }\textbf {\bibinfo
			{volume} {21}},\ \bibinfo {pages} {97} (\bibinfo {year} {2014})}\BibitemShut
	{NoStop}%
	\bibitem [{\citenamefont {Eggeling}(2015)}]{eggeling2015super}%
	\BibitemOpen
	\bibfield  {author} {\bibinfo {author} {\bibfnamefont {C.}~\bibnamefont
			{Eggeling}},\ }\href@noop {} {\bibfield  {journal} {\bibinfo  {journal}
			{Essays in biochemistry}\ }\textbf {\bibinfo {volume} {57}},\ \bibinfo
		{pages} {69} (\bibinfo {year} {2015})}\BibitemShut {NoStop}%
	\bibitem [{\citenamefont {Ortega-Arroyo}\ and\ \citenamefont
		{Kukura}(2012)}]{ortega2012interferometric}%
	\BibitemOpen
	\bibfield  {author} {\bibinfo {author} {\bibfnamefont {J.}~\bibnamefont
			{Ortega-Arroyo}}\ and\ \bibinfo {author} {\bibfnamefont {P.}~\bibnamefont
			{Kukura}},\ }\href@noop {} {\bibfield  {journal} {\bibinfo  {journal}
			{Physical Chemistry Chemical Physics}\ }\textbf {\bibinfo {volume} {14}},\
		\bibinfo {pages} {15625} (\bibinfo {year} {2012})}\BibitemShut {NoStop}%
	\bibitem [{\citenamefont {Kusumi}\ \emph {et~al.}(2005)\citenamefont {Kusumi},
		\citenamefont {Nakada}, \citenamefont {Ritchie}, \citenamefont {Murase},
		\citenamefont {Suzuki}, \citenamefont {Murakoshi}, \citenamefont {Kasai},
		\citenamefont {Kondo},\ and\ \citenamefont {Fujiwara}}]{kusumi2005paradigm}%
	\BibitemOpen
	\bibfield  {author} {\bibinfo {author} {\bibfnamefont {A.}~\bibnamefont
			{Kusumi}}, \bibinfo {author} {\bibfnamefont {C.}~\bibnamefont {Nakada}},
		\bibinfo {author} {\bibfnamefont {K.}~\bibnamefont {Ritchie}}, \bibinfo
		{author} {\bibfnamefont {K.}~\bibnamefont {Murase}}, \bibinfo {author}
		{\bibfnamefont {K.}~\bibnamefont {Suzuki}}, \bibinfo {author} {\bibfnamefont
			{H.}~\bibnamefont {Murakoshi}}, \bibinfo {author} {\bibfnamefont {R.~S.}\
			\bibnamefont {Kasai}}, \bibinfo {author} {\bibfnamefont {J.}~\bibnamefont
			{Kondo}}, \ and\ \bibinfo {author} {\bibfnamefont {T.}~\bibnamefont
			{Fujiwara}},\ }\href@noop {} {\bibfield  {journal} {\bibinfo  {journal}
			{Annu. Rev. Biophys. Biomol. Struct.}\ }\textbf {\bibinfo {volume} {34}},\
		\bibinfo {pages} {351} (\bibinfo {year} {2005})}\BibitemShut {NoStop}%
	\bibitem [{\citenamefont {Chang}\ and\ \citenamefont
		{Rosenthal}(2012)}]{chang2012visualization}%
	\BibitemOpen
	\bibfield  {author} {\bibinfo {author} {\bibfnamefont {J.~C.}\ \bibnamefont
			{Chang}}\ and\ \bibinfo {author} {\bibfnamefont {S.~J.}\ \bibnamefont
			{Rosenthal}},\ }\href@noop {} {\bibfield  {journal} {\bibinfo  {journal} {ACS
				chemical neuroscience}\ }\textbf {\bibinfo {volume} {3}},\ \bibinfo {pages}
		{737} (\bibinfo {year} {2012})}\BibitemShut {NoStop}%
	\bibitem [{\citenamefont {Suzuki}(2016)}]{suzuki2016single}%
	\BibitemOpen
	\bibfield  {author} {\bibinfo {author} {\bibfnamefont {K.~G.}\ \bibnamefont
			{Suzuki}},\ }in\ \href@noop {} {\emph {\bibinfo {booktitle} {Lipid Signaling
				Protocols}}}\ (\bibinfo  {publisher} {Springer},\ \bibinfo {year} {2016})\
	pp.\ \bibinfo {pages} {229--238}\BibitemShut {NoStop}%
	\bibitem [{\citenamefont {de~Almeida}\ \emph {et~al.}(2005)\citenamefont
		{de~Almeida}, \citenamefont {Loura}, \citenamefont {Fedorov},\ and\
		\citenamefont {Prieto}}]{de2005lipid}%
	\BibitemOpen
	\bibfield  {author} {\bibinfo {author} {\bibfnamefont {R.~F.}\ \bibnamefont
			{de~Almeida}}, \bibinfo {author} {\bibfnamefont {L.~M.}\ \bibnamefont
			{Loura}}, \bibinfo {author} {\bibfnamefont {A.}~\bibnamefont {Fedorov}}, \
		and\ \bibinfo {author} {\bibfnamefont {M.}~\bibnamefont {Prieto}},\
	}\href@noop {} {\bibfield  {journal} {\bibinfo  {journal} {Journal of
				molecular biology}\ }\textbf {\bibinfo {volume} {346}},\ \bibinfo {pages}
		{1109} (\bibinfo {year} {2005})}\BibitemShut {NoStop}%
	\bibitem [{\citenamefont {Fijałkowski}(2006)}]{wikimedia}%
	\BibitemOpen
	\bibfield  {author} {\bibinfo {author} {\bibfnamefont {A.~J.}\ \bibnamefont
			{Fijałkowski}},\ }\href
	{https://en.wikipedia.org/wiki/File:Lipid_raft_organisation_scheme.svg}
	{\enquote {\bibinfo {title} {Lipid raft organisation scheme},}\ } (\bibinfo
	{year} {2006})\BibitemShut {NoStop}%
	\bibitem [{\citenamefont {Simons}\ and\ \citenamefont
		{Vaz}(2004)}]{simons2004model}%
	\BibitemOpen
	\bibfield  {author} {\bibinfo {author} {\bibfnamefont {K.}~\bibnamefont
			{Simons}}\ and\ \bibinfo {author} {\bibfnamefont {W.~L.}\ \bibnamefont
			{Vaz}},\ }\href@noop {} {\bibfield  {journal} {\bibinfo  {journal} {Annu.
				Rev. Biophys. Biomol. Struct.}\ }\textbf {\bibinfo {volume} {33}},\ \bibinfo
		{pages} {269} (\bibinfo {year} {2004})}\BibitemShut {NoStop}%
	\bibitem [{\citenamefont {Simons}\ and\ \citenamefont
		{Gerl}(2010)}]{simons2010revitalizing}%
	\BibitemOpen
	\bibfield  {author} {\bibinfo {author} {\bibfnamefont {K.}~\bibnamefont
			{Simons}}\ and\ \bibinfo {author} {\bibfnamefont {M.~J.}\ \bibnamefont
			{Gerl}},\ }\href@noop {} {\bibfield  {journal} {\bibinfo  {journal} {Nature
				reviews Molecular cell biology}\ }\textbf {\bibinfo {volume} {11}},\ \bibinfo
		{pages} {nrm2977} (\bibinfo {year} {2010})}\BibitemShut {NoStop}%
	\bibitem [{\citenamefont {Tamm}\ and\ \citenamefont
		{McConnell}(1985)}]{tamm1985supported}%
	\BibitemOpen
	\bibfield  {author} {\bibinfo {author} {\bibfnamefont {L.~K.}\ \bibnamefont
			{Tamm}}\ and\ \bibinfo {author} {\bibfnamefont {H.~M.}\ \bibnamefont
			{McConnell}},\ }\href@noop {} {\bibfield  {journal} {\bibinfo  {journal}
			{Biophysical journal}\ }\textbf {\bibinfo {volume} {47}},\ \bibinfo {pages}
		{105} (\bibinfo {year} {1985})}\BibitemShut {NoStop}%
	\bibitem [{\citenamefont {Veatch}\ and\ \citenamefont
		{Keller}(2003)}]{veatch2003separation}%
	\BibitemOpen
	\bibfield  {author} {\bibinfo {author} {\bibfnamefont {S.~L.}\ \bibnamefont
			{Veatch}}\ and\ \bibinfo {author} {\bibfnamefont {S.~L.}\ \bibnamefont
			{Keller}},\ }\href@noop {} {\bibfield  {journal} {\bibinfo  {journal}
			{Biophysical journal}\ }\textbf {\bibinfo {volume} {85}},\ \bibinfo {pages}
		{3074} (\bibinfo {year} {2003})}\BibitemShut {NoStop}%
	\bibitem [{\citenamefont {Kahya}\ \emph {et~al.}(2005)\citenamefont {Kahya},
		\citenamefont {Brown},\ and\ \citenamefont {Schwille}}]{kahya2005raft}%
	\BibitemOpen
	\bibfield  {author} {\bibinfo {author} {\bibfnamefont {N.}~\bibnamefont
			{Kahya}}, \bibinfo {author} {\bibfnamefont {D.~A.}\ \bibnamefont {Brown}}, \
		and\ \bibinfo {author} {\bibfnamefont {P.}~\bibnamefont {Schwille}},\
	}\href@noop {} {\bibfield  {journal} {\bibinfo  {journal} {Biochemistry}\
		}\textbf {\bibinfo {volume} {44}},\ \bibinfo {pages} {7479} (\bibinfo {year}
		{2005})}\BibitemShut {NoStop}%
	\bibitem [{\citenamefont {Levental}\ \emph {et~al.}(2009)\citenamefont
		{Levental}, \citenamefont {Byfield}, \citenamefont {Chowdhury}, \citenamefont
		{Gai}, \citenamefont {Baumgart},\ and\ \citenamefont
		{Janmey}}]{levental2009cholesterol}%
	\BibitemOpen
	\bibfield  {author} {\bibinfo {author} {\bibfnamefont {I.}~\bibnamefont
			{Levental}}, \bibinfo {author} {\bibfnamefont {F.~J.}\ \bibnamefont
			{Byfield}}, \bibinfo {author} {\bibfnamefont {P.}~\bibnamefont {Chowdhury}},
		\bibinfo {author} {\bibfnamefont {F.}~\bibnamefont {Gai}}, \bibinfo {author}
		{\bibfnamefont {T.}~\bibnamefont {Baumgart}}, \ and\ \bibinfo {author}
		{\bibfnamefont {P.~A.}\ \bibnamefont {Janmey}},\ }\href@noop {} {\bibfield
		{journal} {\bibinfo  {journal} {Biochemical Journal}\ }\textbf {\bibinfo
			{volume} {424}},\ \bibinfo {pages} {163} (\bibinfo {year}
		{2009})}\BibitemShut {NoStop}%
	\bibitem [{\citenamefont {Hancock}(2006)}]{Hancock2006}%
	\BibitemOpen
	\bibfield  {author} {\bibinfo {author} {\bibfnamefont {J.~F.}\ \bibnamefont
			{Hancock}},\ }\href {\doibase 10.1038/nrm1925} {\bibfield  {journal}
		{\bibinfo  {journal} {Nature Reviews Molecular Cell Biology}\ }\textbf
		{\bibinfo {volume} {7}},\ \bibinfo {pages} {456} (\bibinfo {year}
		{2006})}\BibitemShut {NoStop}%
	\bibitem [{\citenamefont {Levental}\ \emph {et~al.}(2011)\citenamefont
		{Levental}, \citenamefont {Grzybek},\ and\ \citenamefont
		{Simons}}]{levental2011raft}%
	\BibitemOpen
	\bibfield  {author} {\bibinfo {author} {\bibfnamefont {I.}~\bibnamefont
			{Levental}}, \bibinfo {author} {\bibfnamefont {M.}~\bibnamefont {Grzybek}}, \
		and\ \bibinfo {author} {\bibfnamefont {K.}~\bibnamefont {Simons}},\
	}\href@noop {} {\bibfield  {journal} {\bibinfo  {journal} {Proceedings of the
				National Academy of Sciences}\ }\textbf {\bibinfo {volume} {108}},\ \bibinfo
		{pages} {11411} (\bibinfo {year} {2011})}\BibitemShut {NoStop}%
	\bibitem [{\citenamefont {Tian}\ and\ \citenamefont
		{Baumgart}(2009)}]{tian2009sorting}%
	\BibitemOpen
	\bibfield  {author} {\bibinfo {author} {\bibfnamefont {A.}~\bibnamefont
			{Tian}}\ and\ \bibinfo {author} {\bibfnamefont {T.}~\bibnamefont
			{Baumgart}},\ }\href@noop {} {\bibfield  {journal} {\bibinfo  {journal}
			{Biophysical journal}\ }\textbf {\bibinfo {volume} {96}},\ \bibinfo {pages}
		{2676} (\bibinfo {year} {2009})}\BibitemShut {NoStop}%
	\bibitem [{\citenamefont {Heinrich}\ \emph {et~al.}(2010)\citenamefont
		{Heinrich}, \citenamefont {Tian}, \citenamefont {Esposito},\ and\
		\citenamefont {Baumgart}}]{heinrich2010dynamic}%
	\BibitemOpen
	\bibfield  {author} {\bibinfo {author} {\bibfnamefont {M.}~\bibnamefont
			{Heinrich}}, \bibinfo {author} {\bibfnamefont {A.}~\bibnamefont {Tian}},
		\bibinfo {author} {\bibfnamefont {C.}~\bibnamefont {Esposito}}, \ and\
		\bibinfo {author} {\bibfnamefont {T.}~\bibnamefont {Baumgart}},\ }\href@noop
	{} {\bibfield  {journal} {\bibinfo  {journal} {Proceedings of the National
				Academy of Sciences}\ }\textbf {\bibinfo {volume} {107}},\ \bibinfo {pages}
		{7208} (\bibinfo {year} {2010})}\BibitemShut {NoStop}%
	\bibitem [{\citenamefont {Parthasarathy}\ \emph {et~al.}(2006)\citenamefont
		{Parthasarathy}, \citenamefont {Yu},\ and\ \citenamefont
		{Groves}}]{parthasarathy2006curvature}%
	\BibitemOpen
	\bibfield  {author} {\bibinfo {author} {\bibfnamefont {R.}~\bibnamefont
			{Parthasarathy}}, \bibinfo {author} {\bibfnamefont {C.-H.}\ \bibnamefont
			{Yu}}, \ and\ \bibinfo {author} {\bibfnamefont {J.~T.}\ \bibnamefont
			{Groves}},\ }\href@noop {} {\bibfield  {journal} {\bibinfo  {journal}
			{Langmuir}\ }\textbf {\bibinfo {volume} {22}},\ \bibinfo {pages} {5095}
		(\bibinfo {year} {2006})}\BibitemShut {NoStop}%
	\bibitem [{\citenamefont {Cooke}\ and\ \citenamefont
		{Deserno}(2006)}]{cooke2006coupling}%
	\BibitemOpen
	\bibfield  {author} {\bibinfo {author} {\bibfnamefont {I.~R.}\ \bibnamefont
			{Cooke}}\ and\ \bibinfo {author} {\bibfnamefont {M.}~\bibnamefont
			{Deserno}},\ }\href@noop {} {\bibfield  {journal} {\bibinfo  {journal}
			{Biophysical journal}\ }\textbf {\bibinfo {volume} {91}},\ \bibinfo {pages}
		{487} (\bibinfo {year} {2006})}\BibitemShut {NoStop}%
	\bibitem [{\citenamefont {Risselada}\ and\ \citenamefont
		{Marrink}(2009)}]{risselada2009curvature}%
	\BibitemOpen
	\bibfield  {author} {\bibinfo {author} {\bibfnamefont {H.~J.}\ \bibnamefont
			{Risselada}}\ and\ \bibinfo {author} {\bibfnamefont {S.~J.}\ \bibnamefont
			{Marrink}},\ }\href@noop {} {\bibfield  {journal} {\bibinfo  {journal}
			{Physical Chemistry Chemical Physics}\ }\textbf {\bibinfo {volume} {11}},\
		\bibinfo {pages} {2056} (\bibinfo {year} {2009})}\BibitemShut {NoStop}%
	\bibitem [{\citenamefont {R{\'o}{\.z}ycki}\ \emph {et~al.}(2008)\citenamefont
		{R{\'o}{\.z}ycki}, \citenamefont {Weikl},\ and\ \citenamefont
		{Lipowsky}}]{rozycki2008stable}%
	\BibitemOpen
	\bibfield  {author} {\bibinfo {author} {\bibfnamefont {B.}~\bibnamefont
			{R{\'o}{\.z}ycki}}, \bibinfo {author} {\bibfnamefont {T.~R.}\ \bibnamefont
			{Weikl}}, \ and\ \bibinfo {author} {\bibfnamefont {R.}~\bibnamefont
			{Lipowsky}},\ }\href@noop {} {\bibfield  {journal} {\bibinfo  {journal}
			{Physical review letters}\ }\textbf {\bibinfo {volume} {100}},\ \bibinfo
		{pages} {098103} (\bibinfo {year} {2008})}\BibitemShut {NoStop}%
	\bibitem [{\citenamefont {Honerkamp-Smith}\ \emph {et~al.}(2008)\citenamefont
		{Honerkamp-Smith}, \citenamefont {Cicuta}, \citenamefont {Collins},
		\citenamefont {Veatch}, \citenamefont {den Nijs}, \citenamefont {Schick},\
		and\ \citenamefont {Keller}}]{Honerkamp-Smith2008}%
	\BibitemOpen
	\bibfield  {author} {\bibinfo {author} {\bibfnamefont {A.~R.}\ \bibnamefont
			{Honerkamp-Smith}}, \bibinfo {author} {\bibfnamefont {P.}~\bibnamefont
			{Cicuta}}, \bibinfo {author} {\bibfnamefont {M.~D.}\ \bibnamefont {Collins}},
		\bibinfo {author} {\bibfnamefont {S.~L.}\ \bibnamefont {Veatch}}, \bibinfo
		{author} {\bibfnamefont {M.}~\bibnamefont {den Nijs}}, \bibinfo {author}
		{\bibfnamefont {M.}~\bibnamefont {Schick}}, \ and\ \bibinfo {author}
		{\bibfnamefont {S.~L.}\ \bibnamefont {Keller}},\ }\href {\doibase
		10.1529/biophysj.107.128421} {\bibfield  {journal} {\bibinfo  {journal}
			{Biophysical journal}\ }\textbf {\bibinfo {volume} {95}},\ \bibinfo {pages}
		{236} (\bibinfo {year} {2008})}\BibitemShut {NoStop}%
	\bibitem [{\citenamefont {Veatch}\ \emph {et~al.}(2008)\citenamefont {Veatch},
		\citenamefont {Cicuta}, \citenamefont {Sengupta}, \citenamefont
		{Honerkamp-Smith}, \citenamefont {Holowka},\ and\ \citenamefont
		{Baird}}]{veatch2008critical}%
	\BibitemOpen
	\bibfield  {author} {\bibinfo {author} {\bibfnamefont {S.~L.}\ \bibnamefont
			{Veatch}}, \bibinfo {author} {\bibfnamefont {P.}~\bibnamefont {Cicuta}},
		\bibinfo {author} {\bibfnamefont {P.}~\bibnamefont {Sengupta}}, \bibinfo
		{author} {\bibfnamefont {A.}~\bibnamefont {Honerkamp-Smith}}, \bibinfo
		{author} {\bibfnamefont {D.}~\bibnamefont {Holowka}}, \ and\ \bibinfo
		{author} {\bibfnamefont {B.}~\bibnamefont {Baird}},\ }\href@noop {}
	{\bibfield  {journal} {\bibinfo  {journal} {ACS chemical biology}\ }\textbf
		{\bibinfo {volume} {3}},\ \bibinfo {pages} {287} (\bibinfo {year}
		{2008})}\BibitemShut {NoStop}%
	\bibitem [{\citenamefont {Honerkamp-Smith}\ \emph {et~al.}(2009)\citenamefont
		{Honerkamp-Smith}, \citenamefont {Veatch},\ and\ \citenamefont
		{Keller}}]{Honerkamp-Smith2009}%
	\BibitemOpen
	\bibfield  {author} {\bibinfo {author} {\bibfnamefont {A.~R.}\ \bibnamefont
			{Honerkamp-Smith}}, \bibinfo {author} {\bibfnamefont {S.~L.}\ \bibnamefont
			{Veatch}}, \ and\ \bibinfo {author} {\bibfnamefont {S.~L.}\ \bibnamefont
			{Keller}},\ }\href {\doibase 10.1016/j.bbamem.2008.09.010} {\bibfield
		{journal} {\bibinfo  {journal} {Biochimica et Biophysica Acta -
				Biomembranes}\ }\textbf {\bibinfo {volume} {1788}},\ \bibinfo {pages} {53}
		(\bibinfo {year} {2009})}\BibitemShut {NoStop}%
	\bibitem [{\citenamefont {Machta}\ \emph {et~al.}(2012)\citenamefont {Machta},
		\citenamefont {Veatch},\ and\ \citenamefont {Sethna}}]{machta2012critical}%
	\BibitemOpen
	\bibfield  {author} {\bibinfo {author} {\bibfnamefont {B.~B.}\ \bibnamefont
			{Machta}}, \bibinfo {author} {\bibfnamefont {S.~L.}\ \bibnamefont {Veatch}},
		\ and\ \bibinfo {author} {\bibfnamefont {J.~P.}\ \bibnamefont {Sethna}},\
	}\href@noop {} {\bibfield  {journal} {\bibinfo  {journal} {Physical review
				letters}\ }\textbf {\bibinfo {volume} {109}},\ \bibinfo {pages} {138101}
		(\bibinfo {year} {2012})}\BibitemShut {NoStop}%
	\bibitem [{\citenamefont {Honerkamp-Smith}\ \emph {et~al.}(2012)\citenamefont
		{Honerkamp-Smith}, \citenamefont {MacHta},\ and\ \citenamefont
		{Keller}}]{Honerkamp-Smith2012}%
	\BibitemOpen
	\bibfield  {author} {\bibinfo {author} {\bibfnamefont {A.~R.}\ \bibnamefont
			{Honerkamp-Smith}}, \bibinfo {author} {\bibfnamefont {B.~B.}\ \bibnamefont
			{MacHta}}, \ and\ \bibinfo {author} {\bibfnamefont {S.~L.}\ \bibnamefont
			{Keller}},\ }\href {\doibase 10.1103/PhysRevLett.108.265702} {\bibfield
		{journal} {\bibinfo  {journal} {Physical Review Letters}\ }\textbf {\bibinfo
			{volume} {108}},\ \bibinfo {pages} {1} (\bibinfo {year} {2012})}\BibitemShut
	{NoStop}%
	\bibitem [{\citenamefont {Chen}\ \emph {et~al.}(2013)\citenamefont {Chen},
		\citenamefont {Paquette}, \citenamefont {Machta},\ and\ \citenamefont
		{Sethna}}]{Chen2013}%
	\BibitemOpen
	\bibfield  {author} {\bibinfo {author} {\bibfnamefont {Y.-J.}\ \bibnamefont
			{Chen}}, \bibinfo {author} {\bibfnamefont {N.~M.}\ \bibnamefont {Paquette}},
		\bibinfo {author} {\bibfnamefont {B.~B.}\ \bibnamefont {Machta}}, \ and\
		\bibinfo {author} {\bibfnamefont {J.~P.}\ \bibnamefont {Sethna}},\
	}\href@noop {} {\bibfield  {journal} {\bibinfo  {journal} {arXiv}\ }
		(\bibinfo {year} {2013})},\ \Eprint {http://arxiv.org/abs/1307.6899}
	{arXiv:1307.6899} \BibitemShut {NoStop}%
	\bibitem [{\citenamefont {Mitra}\ \emph {et~al.}(2018)\citenamefont {Mitra},
		\citenamefont {Whitehead}, \citenamefont {Holowka}, \citenamefont {Baird},\
		and\ \citenamefont {Sethna}}]{mitra2018computation}%
	\BibitemOpen
	\bibfield  {author} {\bibinfo {author} {\bibfnamefont {E.~D.}\ \bibnamefont
			{Mitra}}, \bibinfo {author} {\bibfnamefont {S.~C.}\ \bibnamefont
			{Whitehead}}, \bibinfo {author} {\bibfnamefont {D.}~\bibnamefont {Holowka}},
		\bibinfo {author} {\bibfnamefont {B.}~\bibnamefont {Baird}}, \ and\ \bibinfo
		{author} {\bibfnamefont {J.~P.}\ \bibnamefont {Sethna}},\ }\href@noop {}
	{\bibfield  {journal} {\bibinfo  {journal} {The Journal of Physical Chemistry
				B}\ }\textbf {\bibinfo {volume} {122}},\ \bibinfo {pages} {3500} (\bibinfo
		{year} {2018})}\BibitemShut {NoStop}%
	\bibitem [{\citenamefont {Gompper}\ and\ \citenamefont
		{Schick}(1990)}]{gompper1990lattice}%
	\BibitemOpen
	\bibfield  {author} {\bibinfo {author} {\bibfnamefont {G.}~\bibnamefont
			{Gompper}}\ and\ \bibinfo {author} {\bibfnamefont {M.}~\bibnamefont
			{Schick}},\ }\href@noop {} {\bibfield  {journal} {\bibinfo  {journal}
			{Physical Review B}\ }\textbf {\bibinfo {volume} {41}},\ \bibinfo {pages}
		{9148} (\bibinfo {year} {1990})}\BibitemShut {NoStop}%
	\bibitem [{\citenamefont {Feigenson}\ and\ \citenamefont
		{Buboltz}(2001)}]{feigenson2001ternary}%
	\BibitemOpen
	\bibfield  {author} {\bibinfo {author} {\bibfnamefont {G.~W.}\ \bibnamefont
			{Feigenson}}\ and\ \bibinfo {author} {\bibfnamefont {J.~T.}\ \bibnamefont
			{Buboltz}},\ }\href@noop {} {\bibfield  {journal} {\bibinfo  {journal}
			{Biophysical journal}\ }\textbf {\bibinfo {volume} {80}},\ \bibinfo {pages}
		{2775} (\bibinfo {year} {2001})}\BibitemShut {NoStop}%
	\bibitem [{\citenamefont {Silvius}(2003)}]{silvius2003fluorescence}%
	\BibitemOpen
	\bibfield  {author} {\bibinfo {author} {\bibfnamefont {J.~R.}\ \bibnamefont
			{Silvius}},\ }\href@noop {} {\bibfield  {journal} {\bibinfo  {journal}
			{Biophysical journal}\ }\textbf {\bibinfo {volume} {85}},\ \bibinfo {pages}
		{1034} (\bibinfo {year} {2003})}\BibitemShut {NoStop}%
	\bibitem [{\citenamefont {Bunge}\ \emph {et~al.}(2008)\citenamefont {Bunge},
		\citenamefont {M{\"u}ller}, \citenamefont {St{\"o}ckl}, \citenamefont
		{Herrmann},\ and\ \citenamefont {Huster}}]{bunge2008characterization}%
	\BibitemOpen
	\bibfield  {author} {\bibinfo {author} {\bibfnamefont {A.}~\bibnamefont
			{Bunge}}, \bibinfo {author} {\bibfnamefont {P.}~\bibnamefont {M{\"u}ller}},
		\bibinfo {author} {\bibfnamefont {M.}~\bibnamefont {St{\"o}ckl}}, \bibinfo
		{author} {\bibfnamefont {A.}~\bibnamefont {Herrmann}}, \ and\ \bibinfo
		{author} {\bibfnamefont {D.}~\bibnamefont {Huster}},\ }\href@noop {}
	{\bibfield  {journal} {\bibinfo  {journal} {Biophysical journal}\ }\textbf
		{\bibinfo {volume} {94}},\ \bibinfo {pages} {2680} (\bibinfo {year}
		{2008})}\BibitemShut {NoStop}%
	\bibitem [{\citenamefont {Chichili}\ and\ \citenamefont
		{Rodgers}(2009)}]{Chichili2009}%
	\BibitemOpen
	\bibfield  {author} {\bibinfo {author} {\bibfnamefont {G.~R.}\ \bibnamefont
			{Chichili}}\ and\ \bibinfo {author} {\bibfnamefont {W.}~\bibnamefont
			{Rodgers}},\ }\href {\doibase 10.1007/s00018-009-0022-6} {\bibfield
		{journal} {\bibinfo  {journal} {Cellular and molecular life sciences}\
		}\textbf {\bibinfo {volume} {66}},\ \bibinfo {pages} {2319} (\bibinfo {year}
		{2009})}\BibitemShut {NoStop}%
	\bibitem [{\citenamefont {Machta}\ \emph {et~al.}(2011)\citenamefont {Machta},
		\citenamefont {Papanikolaou}, \citenamefont {Sethna},\ and\ \citenamefont
		{Veatch}}]{Machta2011}%
	\BibitemOpen
	\bibfield  {author} {\bibinfo {author} {\bibfnamefont {B.~B.}\ \bibnamefont
			{Machta}}, \bibinfo {author} {\bibfnamefont {S.}~\bibnamefont
			{Papanikolaou}}, \bibinfo {author} {\bibfnamefont {J.~P.}\ \bibnamefont
			{Sethna}}, \ and\ \bibinfo {author} {\bibfnamefont {S.~L.}\ \bibnamefont
			{Veatch}},\ }\href {\doibase 10.1016/j.bpj.2011.02.029} {\bibfield  {journal}
		{\bibinfo  {journal} {Biophysical Journal}\ }\textbf {\bibinfo {volume}
			{100}},\ \bibinfo {pages} {1668} (\bibinfo {year} {2011})}\BibitemShut
	{NoStop}%
	\bibitem [{\citenamefont {Head}\ \emph {et~al.}(2014)\citenamefont {Head},
		\citenamefont {Patel},\ and\ \citenamefont {Insel}}]{head2014interaction}%
	\BibitemOpen
	\bibfield  {author} {\bibinfo {author} {\bibfnamefont {B.~P.}\ \bibnamefont
			{Head}}, \bibinfo {author} {\bibfnamefont {H.~H.}\ \bibnamefont {Patel}}, \
		and\ \bibinfo {author} {\bibfnamefont {P.~A.}\ \bibnamefont {Insel}},\
	}\href@noop {} {\bibfield  {journal} {\bibinfo  {journal} {Biochimica et
				Biophysica Acta (BBA)-Biomembranes}\ }\textbf {\bibinfo {volume} {1838}},\
		\bibinfo {pages} {532} (\bibinfo {year} {2014})}\BibitemShut {NoStop}%
	\bibitem [{\citenamefont {Schneider}\ \emph {et~al.}(2017)\citenamefont
		{Schneider}, \citenamefont {Waithe}, \citenamefont {Clausen}, \citenamefont
		{Galiani}, \citenamefont {Koller}, \citenamefont {Ozhan}, \citenamefont
		{Eggeling},\ and\ \citenamefont {Sezgin}}]{schneider2017diffusion}%
	\BibitemOpen
	\bibfield  {author} {\bibinfo {author} {\bibfnamefont {F.}~\bibnamefont
			{Schneider}}, \bibinfo {author} {\bibfnamefont {D.}~\bibnamefont {Waithe}},
		\bibinfo {author} {\bibfnamefont {M.~P.}\ \bibnamefont {Clausen}}, \bibinfo
		{author} {\bibfnamefont {S.}~\bibnamefont {Galiani}}, \bibinfo {author}
		{\bibfnamefont {T.}~\bibnamefont {Koller}}, \bibinfo {author} {\bibfnamefont
			{G.}~\bibnamefont {Ozhan}}, \bibinfo {author} {\bibfnamefont
			{C.}~\bibnamefont {Eggeling}}, \ and\ \bibinfo {author} {\bibfnamefont
			{E.}~\bibnamefont {Sezgin}},\ }\href@noop {} {\bibfield  {journal} {\bibinfo
			{journal} {Molecular biology of the cell}\ }\textbf {\bibinfo {volume}
			{28}},\ \bibinfo {pages} {1507} (\bibinfo {year} {2017})}\BibitemShut
	{NoStop}%
	\bibitem [{\citenamefont {Cooke}\ \emph {et~al.}(2005)\citenamefont {Cooke},
		\citenamefont {Kremer},\ and\ \citenamefont {Deserno}}]{Cooke2005}%
	\BibitemOpen
	\bibfield  {author} {\bibinfo {author} {\bibfnamefont {I.~R.}\ \bibnamefont
			{Cooke}}, \bibinfo {author} {\bibfnamefont {K.}~\bibnamefont {Kremer}}, \
		and\ \bibinfo {author} {\bibfnamefont {M.}~\bibnamefont {Deserno}},\
	}\href@noop {} {\bibfield  {journal} {\bibinfo  {journal} {Physical Review
				E}\ }\textbf {\bibinfo {volume} {72}},\ \bibinfo {pages} {011506} (\bibinfo
		{year} {2005})}\BibitemShut {NoStop}%
	\bibitem [{\citenamefont {Komura}\ \emph {et~al.}(2016)\citenamefont {Komura},
		\citenamefont {Suzuki}, \citenamefont {Ando}, \citenamefont {Konishi},
		\citenamefont {Koikeda}, \citenamefont {Imamura}, \citenamefont {Chadda},
		\citenamefont {Fujiwara}, \citenamefont {Tsuboi}, \citenamefont {Sheng},
		\citenamefont {Cho}, \citenamefont {Furukawa}, \citenamefont {Furukawa},
		\citenamefont {Yamauchi}, \citenamefont {Ishida}, \citenamefont {Kusumi},\
		and\ \citenamefont {Kiso}}]{Komura2016}%
	\BibitemOpen
	\bibfield  {author} {\bibinfo {author} {\bibfnamefont {N.}~\bibnamefont
			{Komura}}, \bibinfo {author} {\bibfnamefont {K.~G.~N.}\ \bibnamefont
			{Suzuki}}, \bibinfo {author} {\bibfnamefont {H.}~\bibnamefont {Ando}},
		\bibinfo {author} {\bibfnamefont {M.}~\bibnamefont {Konishi}}, \bibinfo
		{author} {\bibfnamefont {M.}~\bibnamefont {Koikeda}}, \bibinfo {author}
		{\bibfnamefont {A.}~\bibnamefont {Imamura}}, \bibinfo {author} {\bibfnamefont
			{R.}~\bibnamefont {Chadda}}, \bibinfo {author} {\bibfnamefont {T.~K.}\
			\bibnamefont {Fujiwara}}, \bibinfo {author} {\bibfnamefont {H.}~\bibnamefont
			{Tsuboi}}, \bibinfo {author} {\bibfnamefont {R.}~\bibnamefont {Sheng}},
		\bibinfo {author} {\bibfnamefont {W.}~\bibnamefont {Cho}}, \bibinfo {author}
		{\bibfnamefont {K.}~\bibnamefont {Furukawa}}, \bibinfo {author}
		{\bibfnamefont {K.}~\bibnamefont {Furukawa}}, \bibinfo {author}
		{\bibfnamefont {Y.}~\bibnamefont {Yamauchi}}, \bibinfo {author}
		{\bibfnamefont {H.}~\bibnamefont {Ishida}}, \bibinfo {author} {\bibfnamefont
			{A.}~\bibnamefont {Kusumi}}, \ and\ \bibinfo {author} {\bibfnamefont
			{M.}~\bibnamefont {Kiso}},\ }\href {http://dx.doi.org/10.1038/nchembio.2059}
	{\bibfield  {journal} {\bibinfo  {journal} {Nature Chemical Biology}\
		}\textbf {\bibinfo {volume} {12}},\ \bibinfo {pages} {402 EP } (\bibinfo
		{year} {2016})},\ \bibinfo {note} {article}\BibitemShut {NoStop}%
	\bibitem [{\citenamefont {Suzuki}\ \emph {et~al.}(2018)\citenamefont {Suzuki},
		\citenamefont {Ando}, \citenamefont {Komura}, \citenamefont {Konishi},
		\citenamefont {Imamura}, \citenamefont {Ishida}, \citenamefont {Kiso},
		\citenamefont {Fujiwara},\ and\ \citenamefont {Kusumi}}]{suzuki2018}%
	\BibitemOpen
	\bibfield  {author} {\bibinfo {author} {\bibfnamefont {K.~G.}\ \bibnamefont
			{Suzuki}}, \bibinfo {author} {\bibfnamefont {H.}~\bibnamefont {Ando}},
		\bibinfo {author} {\bibfnamefont {N.}~\bibnamefont {Komura}}, \bibinfo
		{author} {\bibfnamefont {M.}~\bibnamefont {Konishi}}, \bibinfo {author}
		{\bibfnamefont {A.}~\bibnamefont {Imamura}}, \bibinfo {author} {\bibfnamefont
			{H.}~\bibnamefont {Ishida}}, \bibinfo {author} {\bibfnamefont
			{M.}~\bibnamefont {Kiso}}, \bibinfo {author} {\bibfnamefont {T.~K.}\
			\bibnamefont {Fujiwara}}, \ and\ \bibinfo {author} {\bibfnamefont
			{A.}~\bibnamefont {Kusumi}},\ }in\ \href@noop {} {\emph {\bibinfo {booktitle}
			{Methods in enzymology}}},\ Vol.\ \bibinfo {volume} {598}\ (\bibinfo
	{publisher} {Elsevier},\ \bibinfo {year} {2018})\ pp.\ \bibinfo {pages}
	{267--282}\BibitemShut {NoStop}%
	\bibitem [{\citenamefont {Starr}\ \emph {et~al.}(2014)\citenamefont {Starr},
		\citenamefont {Hartmann},\ and\ \citenamefont
		{Douglas}}]{starr2014dynamical}%
	\BibitemOpen
	\bibfield  {author} {\bibinfo {author} {\bibfnamefont {F.~W.}\ \bibnamefont
			{Starr}}, \bibinfo {author} {\bibfnamefont {B.}~\bibnamefont {Hartmann}}, \
		and\ \bibinfo {author} {\bibfnamefont {J.~F.}\ \bibnamefont {Douglas}},\
	}\href@noop {} {\bibfield  {journal} {\bibinfo  {journal} {Soft Matter}\
		}\textbf {\bibinfo {volume} {10}},\ \bibinfo {pages} {3036} (\bibinfo {year}
		{2014})}\BibitemShut {NoStop}%
	\bibitem [{\citenamefont {Andersen}\ \emph {et~al.}(1971)\citenamefont
		{Andersen}, \citenamefont {Weeks},\ and\ \citenamefont
		{Chandler}}]{Andersen1971}%
	\BibitemOpen
	\bibfield  {author} {\bibinfo {author} {\bibfnamefont {H.~C.}\ \bibnamefont
			{Andersen}}, \bibinfo {author} {\bibfnamefont {J.~D.}\ \bibnamefont {Weeks}},
		\ and\ \bibinfo {author} {\bibfnamefont {D.}~\bibnamefont {Chandler}},\
	}\href {\doibase 10.1103/PhysRevA.4.1597} {\bibfield  {journal} {\bibinfo
			{journal} {Physical Review A}\ }\textbf {\bibinfo {volume} {4}},\ \bibinfo
		{pages} {1597} (\bibinfo {year} {1971})}\BibitemShut {NoStop}%
	\bibitem [{\citenamefont {Plimpton}(1995)}]{Plimpton1995}%
	\BibitemOpen
	\bibfield  {author} {\bibinfo {author} {\bibfnamefont {S.}~\bibnamefont
			{Plimpton}},\ }\href {\doibase 10.1006/jcph.1995.1039} {\bibfield  {journal}
		{\bibinfo  {journal} {Journal of Computational Physics}\ }\textbf {\bibinfo
			{volume} {117}},\ \bibinfo {pages} {1} (\bibinfo {year} {1995})}\BibitemShut
	{NoStop}%
	\bibitem [{\citenamefont {Bonfanti}\ \emph {et~al.}(2019)\citenamefont
		{Bonfanti}, \citenamefont {Hoferer}, \citenamefont {Sellerio},\ and\
		\citenamefont {Costantini}}]{github_lammps}%
	\BibitemOpen
	\bibfield  {author} {\bibinfo {author} {\bibfnamefont {S.}~\bibnamefont
			{Bonfanti}}, \bibinfo {author} {\bibfnamefont {M.}~\bibnamefont {Hoferer}},
		\bibinfo {author} {\bibfnamefont {A.~L.}\ \bibnamefont {Sellerio}}, \ and\
		\bibinfo {author} {\bibfnamefont {G.}~\bibnamefont {Costantini}},\ }\href
	{https://github.com/ComplexityBiosystems/Flat_Membrane_with_Protein_MD_Simulations}
	{\enquote {\bibinfo {title} {Simulation code of a flat lipid membrane with
				additional protein in lammps},}\ }\bibinfo {howpublished}
	{\url{https://github.com/ComplexityBiosystems/Flat_Membrane_with_Protein_MD_Simulations}}
	(\bibinfo {year} {2018 (accessed on September 9, 2019})\BibitemShut {NoStop}%
	\bibitem [{\citenamefont {Cooke}\ and\ \citenamefont
		{Deserno}(2005)}]{cooke2005solvent}%
	\BibitemOpen
	\bibfield  {author} {\bibinfo {author} {\bibfnamefont {I.~R.}\ \bibnamefont
			{Cooke}}\ and\ \bibinfo {author} {\bibfnamefont {M.}~\bibnamefont
			{Deserno}},\ }\href@noop {} {\bibfield  {journal} {\bibinfo  {journal} {The
				Journal of chemical physics}\ }\textbf {\bibinfo {volume} {123}},\ \bibinfo
		{pages} {224710} (\bibinfo {year} {2005})}\BibitemShut {NoStop}%
	\bibitem [{\citenamefont {Tristram-Nagle}\ and\ \citenamefont
		{Nagle}(2004)}]{tristram2004lipid}%
	\BibitemOpen
	\bibfield  {author} {\bibinfo {author} {\bibfnamefont {S.}~\bibnamefont
			{Tristram-Nagle}}\ and\ \bibinfo {author} {\bibfnamefont {J.~F.}\
			\bibnamefont {Nagle}},\ }\href@noop {} {\bibfield  {journal} {\bibinfo
			{journal} {Chemistry and physics of lipids}\ }\textbf {\bibinfo {volume}
			{127}},\ \bibinfo {pages} {3} (\bibinfo {year} {2004})}\BibitemShut {NoStop}%
	\bibitem [{\citenamefont {Klingelhoefer}\ \emph {et~al.}(2009)\citenamefont
		{Klingelhoefer}, \citenamefont {Carpenter},\ and\ \citenamefont
		{Sansom}}]{klingelhoefer2009peptide}%
	\BibitemOpen
	\bibfield  {author} {\bibinfo {author} {\bibfnamefont {J.~W.}\ \bibnamefont
			{Klingelhoefer}}, \bibinfo {author} {\bibfnamefont {T.}~\bibnamefont
			{Carpenter}}, \ and\ \bibinfo {author} {\bibfnamefont {M.~S.}\ \bibnamefont
			{Sansom}},\ }\href@noop {} {\bibfield  {journal} {\bibinfo  {journal}
			{Biophysical journal}\ }\textbf {\bibinfo {volume} {96}},\ \bibinfo {pages}
		{3519} (\bibinfo {year} {2009})}\BibitemShut {NoStop}%
	\bibitem [{\citenamefont {Hoferer}(2019{\natexlab{a}})}]{S1_MD_video}%
	\BibitemOpen
	\bibfield  {author} {\bibinfo {author} {\bibfnamefont {M.}~\bibnamefont
			{Hoferer}},\ }\href {https://youtu.be/1IW5CkmEDJQ} {\enquote {\bibinfo
			{title} {Coarse-grained molecular dynamics model simulation video},}\
	}\bibinfo {howpublished} {YouTube} (\bibinfo {year}
	{2019}{\natexlab{a}})\BibitemShut {NoStop}%
	\bibitem [{\citenamefont {Hoferer}(2019{\natexlab{b}})}]{S2_lattice_video}%
	\BibitemOpen
	\bibfield  {author} {\bibinfo {author} {\bibfnamefont {M.}~\bibnamefont
			{Hoferer}},\ }\href {https://youtu.be/9HVMokl4n2As} {\enquote {\bibinfo
			{title} {Ising-like 2{D} lattice model simulation video},}\ }\bibinfo
	{howpublished} {YouTube} (\bibinfo {year} {2019}{\natexlab{b}})\BibitemShut
	{NoStop}%
	\bibitem [{\citenamefont {Hoferer}(2017)}]{github_lattice_model}%
	\BibitemOpen
	\bibfield  {author} {\bibinfo {author} {\bibfnamefont {M.}~\bibnamefont
			{Hoferer}},\ }\href {https://github.com/moritzhoferer/lattice_membrane}
	{\enquote {\bibinfo {title} {Lattice model for simulations of lipid and
				protein interactions in a flat membrane},}\ }\bibinfo {howpublished} {GitHub}
	(\bibinfo {year} {2017})\BibitemShut {NoStop}%
	\bibitem [{\citenamefont {Krapivsky}\ \emph {et~al.}(2010)\citenamefont
		{Krapivsky}, \citenamefont {Redner},\ and\ \citenamefont
		{Ben-Naim}}]{krapivsky2010kinetic}%
	\BibitemOpen
	\bibfield  {author} {\bibinfo {author} {\bibfnamefont {P.}~\bibnamefont
			{Krapivsky}}, \bibinfo {author} {\bibfnamefont {S.}~\bibnamefont {Redner}}, \
		and\ \bibinfo {author} {\bibfnamefont {E.}~\bibnamefont {Ben-Naim}},\ }\href
	{https://books.google.it/books?id=cc3pApnX3kYC} {\emph {\bibinfo {title} {A
				Kinetic View of Statistical Physics}}}\ (\bibinfo  {publisher} {Cambridge
		University Press},\ \bibinfo {year} {2010})\BibitemShut {NoStop}%
	\bibitem [{\citenamefont {Kawasaki}(1966)}]{Kawasaki1965}%
	\BibitemOpen
	\bibfield  {author} {\bibinfo {author} {\bibfnamefont {K.}~\bibnamefont
			{Kawasaki}},\ }\href {\doibase 10.1103/PhysRev.145.224} {\bibfield  {journal}
		{\bibinfo  {journal} {Physical Review Letters}\ }\textbf {\bibinfo {volume}
			{145}},\ \bibinfo {pages} {224} (\bibinfo {year} {1966})}\BibitemShut
	{NoStop}%
	\bibitem [{\citenamefont {Glauber}(1963)}]{Glauber1963}%
	\BibitemOpen
	\bibfield  {author} {\bibinfo {author} {\bibfnamefont {R.~J.}\ \bibnamefont
			{Glauber}},\ }\href {\doibase 10.1063/1.1703954} {\bibfield  {journal}
		{\bibinfo  {journal} {Journal of Mathematical Physics}\ }\textbf {\bibinfo
			{volume} {4}},\ \bibinfo {pages} {294} (\bibinfo {year} {1963})}\BibitemShut
	{NoStop}%
	\bibitem [{\citenamefont {Baxter}(1982)}]{Baxter_1982}%
	\BibitemOpen
	\bibfield  {author} {\bibinfo {author} {\bibfnamefont {R.~J.}\ \bibnamefont
			{Baxter}},\ }\href {http://opac.inria.fr/record=b1128399} {\emph {\bibinfo
			{title} {Exactly solved models in statistical mechanics}}}\ (\bibinfo
	{publisher} {Academic Press},\ \bibinfo {address} {London},\ \bibinfo {year}
	{1982})\BibitemShut {NoStop}%
	\bibitem [{\citenamefont {Plischke}\ and\ \citenamefont
		{Bergersen}(2006)}]{Plischke2006}%
	\BibitemOpen
	\bibfield  {author} {\bibinfo {author} {\bibfnamefont {M.}~\bibnamefont
			{Plischke}}\ and\ \bibinfo {author} {\bibfnamefont {B.}~\bibnamefont
			{Bergersen}},\ }\href {https://books.google.it/books?id=KYu7igYEkhwC} {\emph
		{\bibinfo {title} {Equilibrium Statistical Physics}}}\ (\bibinfo  {publisher}
	{World Scientific},\ \bibinfo {year} {2006})\BibitemShut {NoStop}%
	\bibitem [{\citenamefont {Metropolis}\ \emph {et~al.}(1953)\citenamefont
		{Metropolis}, \citenamefont {Rosenbluth}, \citenamefont {Rosenbluth},
		\citenamefont {Teller},\ and\ \citenamefont {Teller}}]{Metropolis1953}%
	\BibitemOpen
	\bibfield  {author} {\bibinfo {author} {\bibfnamefont {N.}~\bibnamefont
			{Metropolis}}, \bibinfo {author} {\bibfnamefont {A.~W.}\ \bibnamefont
			{Rosenbluth}}, \bibinfo {author} {\bibfnamefont {M.~N.}\ \bibnamefont
			{Rosenbluth}}, \bibinfo {author} {\bibfnamefont {A.~H.}\ \bibnamefont
			{Teller}}, \ and\ \bibinfo {author} {\bibfnamefont {E.}~\bibnamefont
			{Teller}},\ }\href {\doibase http://dx.doi.org/10.1063/1.1699114} {\bibfield
		{journal} {\bibinfo  {journal} {Journal Chemical Physics}\ }\textbf {\bibinfo
			{volume} {21}},\ \bibinfo {pages} {1087} (\bibinfo {year}
		{1953})}\BibitemShut {NoStop}%
	\bibitem [{\citenamefont {Hoshen}\ and\ \citenamefont
		{Kopelman}(1976)}]{Hoshen1976}%
	\BibitemOpen
	\bibfield  {author} {\bibinfo {author} {\bibfnamefont {J.}~\bibnamefont
			{Hoshen}}\ and\ \bibinfo {author} {\bibfnamefont {R.}~\bibnamefont
			{Kopelman}},\ }\href {\doibase 10.1103/PhysRevB.14.3438} {\bibfield
		{journal} {\bibinfo  {journal} {Physical Review B}\ }\textbf {\bibinfo
			{volume} {14}},\ \bibinfo {pages} {3438} (\bibinfo {year}
		{1976})}\BibitemShut {NoStop}%
	\bibitem [{\citenamefont {Hurst}(1951)}]{Hurst1951}%
	\BibitemOpen
	\bibfield  {author} {\bibinfo {author} {\bibfnamefont {H.~E.}\ \bibnamefont
			{Hurst}},\ }\href@noop {} {\bibfield  {journal} {\bibinfo  {journal} {Trans.
				Amer. Soc. Civil Eng.}\ }\textbf {\bibinfo {volume} {116}},\ \bibinfo {pages}
		{770} (\bibinfo {year} {1951})}\BibitemShut {NoStop}%
	\bibitem [{\citenamefont {Mandelbrot}\ and\ \citenamefont
		{Van~Ness}(1968)}]{Mandelbrot1968}%
	\BibitemOpen
	\bibfield  {author} {\bibinfo {author} {\bibfnamefont {B.~B.}\ \bibnamefont
			{Mandelbrot}}\ and\ \bibinfo {author} {\bibfnamefont {J.~W.}\ \bibnamefont
			{Van~Ness}},\ }\href@noop {} {\bibfield  {journal} {\bibinfo  {journal} {SIAM
				review}\ }\textbf {\bibinfo {volume} {10}},\ \bibinfo {pages} {422} (\bibinfo
		{year} {1968})}\BibitemShut {NoStop}%
	\bibitem [{\citenamefont {Metzler}\ and\ \citenamefont
		{Klafter}(2000)}]{Metzler2000}%
	\BibitemOpen
	\bibfield  {author} {\bibinfo {author} {\bibfnamefont {R.}~\bibnamefont
			{Metzler}}\ and\ \bibinfo {author} {\bibfnamefont {J.}~\bibnamefont
			{Klafter}},\ }\href@noop {} {\bibfield  {journal} {\bibinfo  {journal}
			{Physics reports}\ }\textbf {\bibinfo {volume} {339}},\ \bibinfo {pages} {1}
		(\bibinfo {year} {2000})}\BibitemShut {NoStop}%
	\bibitem [{\citenamefont {Taloni}\ \emph {et~al.}(2010)\citenamefont {Taloni},
		\citenamefont {Chechkin},\ and\ \citenamefont {Klafter}}]{Taloni2010}%
	\BibitemOpen
	\bibfield  {author} {\bibinfo {author} {\bibfnamefont {A.}~\bibnamefont
			{Taloni}}, \bibinfo {author} {\bibfnamefont {A.}~\bibnamefont {Chechkin}}, \
		and\ \bibinfo {author} {\bibfnamefont {J.}~\bibnamefont {Klafter}},\
	}\href@noop {} {\bibfield  {journal} {\bibinfo  {journal} {Physical review
				letters}\ }\textbf {\bibinfo {volume} {104}},\ \bibinfo {pages} {160602}
		(\bibinfo {year} {2010})}\BibitemShut {NoStop}%
	\bibitem [{\citenamefont {Lee}(1977)}]{lee1977annular}%
	\BibitemOpen
	\bibfield  {author} {\bibinfo {author} {\bibfnamefont {A.}~\bibnamefont
			{Lee}},\ }\href@noop {} {\bibfield  {journal} {\bibinfo  {journal} {Trends in
				Biochemical Sciences}\ }\textbf {\bibinfo {volume} {2}},\ \bibinfo {pages}
		{231} (\bibinfo {year} {1977})}\BibitemShut {NoStop}%
	\bibitem [{\citenamefont {Watts}\ \emph {et~al.}(1979)\citenamefont {Watts},
		\citenamefont {Volotovski},\ and\ \citenamefont
		{Marsh}}]{watts1979rhodopsin}%
	\BibitemOpen
	\bibfield  {author} {\bibinfo {author} {\bibfnamefont {A.}~\bibnamefont
			{Watts}}, \bibinfo {author} {\bibfnamefont {I.~D.}\ \bibnamefont
			{Volotovski}}, \ and\ \bibinfo {author} {\bibfnamefont {D.}~\bibnamefont
			{Marsh}},\ }\href@noop {} {\bibfield  {journal} {\bibinfo  {journal}
			{Biochemistry}\ }\textbf {\bibinfo {volume} {18}},\ \bibinfo {pages} {5006}
		(\bibinfo {year} {1979})}\BibitemShut {NoStop}%
	\bibitem [{\citenamefont {Watts}\ \emph {et~al.}(1981)\citenamefont {Watts},
		\citenamefont {Davoust}, \citenamefont {Marsh},\ and\ \citenamefont
		{Devaux}}]{watts1981distinct}%
	\BibitemOpen
	\bibfield  {author} {\bibinfo {author} {\bibfnamefont {A.}~\bibnamefont
			{Watts}}, \bibinfo {author} {\bibfnamefont {J.}~\bibnamefont {Davoust}},
		\bibinfo {author} {\bibfnamefont {D.}~\bibnamefont {Marsh}}, \ and\ \bibinfo
		{author} {\bibfnamefont {P.}~\bibnamefont {Devaux}},\ }\href@noop {}
	{\bibfield  {journal} {\bibinfo  {journal} {Biochimica et Biophysica Acta
				(BBA)-Biomembranes}\ }\textbf {\bibinfo {volume} {643}},\ \bibinfo {pages}
		{673} (\bibinfo {year} {1981})}\BibitemShut {NoStop}%
	\bibitem [{\citenamefont {Davoust}\ and\ \citenamefont
		{Devaux}(1982)}]{davoust1982simulation}%
	\BibitemOpen
	\bibfield  {author} {\bibinfo {author} {\bibfnamefont {J.}~\bibnamefont
			{Davoust}}\ and\ \bibinfo {author} {\bibfnamefont {P.~F.}\ \bibnamefont
			{Devaux}},\ }\href@noop {} {\bibfield  {journal} {\bibinfo  {journal}
			{Journal of Magnetic Resonance (1969)}\ }\textbf {\bibinfo {volume} {48}},\
		\bibinfo {pages} {475} (\bibinfo {year} {1982})}\BibitemShut {NoStop}%
	\bibitem [{\citenamefont {Sternberg}\ \emph {et~al.}(1992)\citenamefont
		{Sternberg}, \citenamefont {L'Hostis}, \citenamefont {Whiteway},\ and\
		\citenamefont {Watts}}]{sternberg1992essential}%
	\BibitemOpen
	\bibfield  {author} {\bibinfo {author} {\bibfnamefont {B.}~\bibnamefont
			{Sternberg}}, \bibinfo {author} {\bibfnamefont {C.}~\bibnamefont {L'Hostis}},
		\bibinfo {author} {\bibfnamefont {C.~A.}\ \bibnamefont {Whiteway}}, \ and\
		\bibinfo {author} {\bibfnamefont {A.}~\bibnamefont {Watts}},\ }\href@noop {}
	{\bibfield  {journal} {\bibinfo  {journal} {Biochimica et Biophysica Acta
				(BBA)-Biomembranes}\ }\textbf {\bibinfo {volume} {1108}},\ \bibinfo {pages}
		{21} (\bibinfo {year} {1992})}\BibitemShut {NoStop}%
\end{thebibliography}
%
\end{document}